\documentclass[12pt]{iopart}

 \usepackage[pdftex]{graphicx}  
\begin{document}

\title[]{The production of gauge bosons pairs $ W ^ {+} W ^ {-} $ associated with 0, 1 and 2 jets in proton-proton collisions at LHC}

\author{K. Djamaa$^{(a)}$, A. Mohamed-Meziani$^{(a,b)}$}

\address{$^{(a)}$ Department of Physics, Faculty of Exact Sciences, University of Bejaia, 06000 Bejaia, Algeria }
\address{$^{(b)}$ Laboratoire de Physique Th\'eorique d'Oran (LPTO), University of Oran1-Ahmed Ben Bella; BP 1524 El M\'\ nouar Oran 31000, Algeria}

\ead{kenza.djamaa@univ-bejaia.dz, abdelkader.mohamedmeziani@univ-bejaia.dz}

\vspace{10pt}
\begin{indented}
\item[]August 2020
\end{indented}

\begin{abstract}
We report in this work  the production of $ W^{+}W^{-}$ pairs  gauge bosons associated with 0, 1 and 2 jets in  proton-proton collisions at LHC with an energy of 14 TeV in the center of mass. These processes are produced at leading-order (LO) and next-to-leading-order (NLO) with QCD corrections in the standard model, using MadGraph$\_$aMC@NLO. For a realistic description of the processes, we match the hard scattering processes with Pythia8 parton showering and hadronization. The obtained events are run through the fast detector simulation, Delphes, which serves to accurately model the ATLAS and CMS detectors and the final state reconstructions that are performed. We analyze the total cross sections according on two cuts in jet transverse momentum  $p_{T,j} > 20$ GeV and $p_{T,j} > 100 $ GeV. We describe the important  numerical aspects of our calculations by presenting transverse momentum and rapidity distributions at partonic fixed order and at parton shower for both W and the $W^{+}W^{-}$ pair.
\end{abstract}

%
%
%
%
%

\section{Introduction}
\label{sec:1}

With the  start of the LHC, operation hard scattering processes became accessible with parton energies up to several TeV and, indeed, jet-jet invariant masses up to 5 TeV have been observed ~\cite{ref:1, ref:2}  in the first round of data taking. 

With the significant increase in integrated luminosity and beam energy that has occurred in recent years, electroweak processes, such as the production of lepton pairs or gauge bosons with invariant masses up to 13 TeV, are accessible. These reactions may on the one hand allow for precise tests of the Standard Model (SM) and on the other hand potential deviations may point to "physics beyond the SM". Indeed, the observation of anomalous couplings of quarks, leptons or gauge bosons might well be the first signal of "New Physics".

Indeed, the first results were presented from searches for the standard model Higgs boson in proton-proton collisions at $\sqrt{s}$ = 7 and 8 TeV in the CMS experiment at the LHC. It used data samples corresponding to integrated luminosities of up to 5.1 $fb^{-1}$  at 7 TeV and 5.3 $fb^{-1}$ at 8 TeV. The search was performed in five decay modes: $\gamma\gamma$, $ZZ$, $W^{+}W^{-}$,$\tau^{+}\tau^{-}$, and bb~\cite{ref:3}. Similarly,  in the ATLAS experiment, integrated luminosities of approximately 4.8 $fb^{-1}$ collected at $\sqrt{s} $= 7 TeV in 2011 and 5.8 $fb^{-1}$ at $\sqrt(s)$ = 8 TeV in 2012. Individual searches in the channels H $\rightarrow  ZZ^{*}  \rightarrow$ 4$l$, H $\rightarrow \gamma\gamma$ and H $\rightarrow WW^{*} \rightarrow e\nu \mu_{\nu}$ in the 8 TeV data were combined with previously published results of searches for H $\rightarrow ZZ^{*}$, $WW^{*}$ , b b and $\tau^{+}\tau^{-}$ in the 7 TeV data and results from improved analyses of the H $\rightarrow ZZ^{*} \rightarrow $ 4$ l$ and H $\rightarrow \gamma\gamma $  channels in the 7 TeV data~\cite{ref:4}. 

The extremely complicated hadron collider environment does not only require sufficiently precise predictions for new physics signals, but also for many complicated background reactions that cannot entirely be measured from data. Among such background processes, several involve three, four, or even more particles in the final state, rendering the necessary next-to-leading-order (NLO) calculations in QCD very complicated. Even at NLO, the calculation of 2 $\rightarrow$  3 (and 2 $\rightarrow$  4) processes is extremely time, so clear priority needs to be established for those processes most needed for the LHC. In the 2005 Les Houches proceedings, such a realistic NLO wishlist was established ~\cite{ref:5,ref:6,ref:7}.

Vector-boson pair production ranks among the most important Standard-Model benchmark processes at the LHC.  Several studies addressed the production of W-boson pairs in hadron collisions, including NLO QCD corrections. In particular, pp $\rightarrow W^{+}W^{-}$  with no jets was studied in Refs ~\cite{ref:8,ref:9,ref:10,ref:11,ref:12},  while the study of the effects of the complete logarithmic electroweak O($\alpha$) corrections on the production of vector-boson pairs WZ, ZZ, and WW at the LHC were performed in ref ~\cite{ref:13}. QCD corrections to $W^{+}W^{-}$ production through gluon fusion was performed in Ref~\cite{ref:14}. 
The $W^{+}W^{-}$  pair production through the leptonic decay channels with Herwig7 is calculated in ~\cite{ref:15}, and at NLO QCD including effective field theory (EFT) in~\cite{ref:16}. The production of a pair of W -bosons in association with one jet including decays to leptons was studied in refs~\cite{ref:17,ref:18}. In both cases, for the choice of the renormalization and factorization scales $\mu$ = $M_{W} $, QCD corrections were found to be significant, of the order of (25 - 50) $\%$. These results further motivate the need to understand the production of $W^{+}W^{-} $ in association with two or even more jets at NLO in QCD. 

The vector bosons pair productions, associated with two jets, pp$\rightarrow$ V V +2jets, are an important background to new Physics and Higgs boson searches. They can be produced through both vector boson fusion (VBF) or production mechanisms without weak bosons in the $t$-channel. NLO QCD corrections to the VBF mechanism have been calculated in ~\cite{ref:19,ref:20}. For the non-VBF case, several calculations have been presented. The production of equal-charge W-bosons, $W^{+}W^{+} jj$, has been computed at NLO accuracy in ~\cite{ref:21,ref:22}. It also has been combined with a parton shower ~\cite{ref:23,ref:24} within the POWHEG framework~\cite{ref:25,ref:26}, and later has been complemented by EW corrections~\cite{ref:23}. First results for NLO QCD corrections to the process $W^{+}W^{-} jj$ at hadron colliders have been presented in ~\cite{ref:27,ref:28} and those performed using the GoSam package for the virtual contributions~\cite{ref:29}, in combination with MadGraph/MadEvent ~\cite{ref:30,ref:31} and MadDipole~\cite{ref:32,ref:33} for the real radiation and the phase space integration. These have included leptonic decays of the W-bosons with full spin correlations.

In this paper we report on the NLO QCD correction  of the $W^{+}W^{-}$, $W^{+}W^{-} j$ and $W^{+} W^{-} jj$ productions both at fixed order and through a Parton Shower analysis. We consider two cuts in the associated jets and we  present a prediction of the total cross-section for these productions at $\sqrt{s} $ = 14 TeV. As well, we present the contribution of the gluon gluon fusion to the total cross section.
These calculations are performed using MadGraph$\_$aMC@NLO ~\cite{ref:34, ref:35}. We use the global data set which includes all the relevant LHC data for which experimental systematic uncertainties are available through the recent NNPDF2.3 PDF sets ~\cite{ref:36}, in the generation of events and calculations of the cross sections as well as the public delphes ATLAS 1604 07773 card analysis~\cite{ref:37} for the reconstruction of the states after the showering. In addition, we also study the W's distributions over the different variables as rapidity and transverse momentum of $W^{+}W^{-}$ pair.

The remainder of the paper is organized as follows. In Section ~\ref{sec:2} we provide details of the NLO total cross section calculations. In Section~\ref{sec:3} we briefly discuss technical aspects of the calculations. In Section~\ref{sec:4} we describe the results. In Section~\ref{sec:5} we present our conclusions. 

\section{NLO total cross sections}
\label{sec:2}

Cross sections at NLO precision are given by

\begin{equation}
\label{eq:1}
\sigma = \sigma^{LO} + \sigma^{NLO}.  
\end{equation}

For point-like initial states, the LO part of $\sigma^{LO}$ is obtained by integrating the exclusive cross section in Borm approximation over the available phase space of the final state particles as bellow:

\begin{equation}
\label{eq:2}
\sigma^{LO} =\int_{m}d^{(4)} \sigma^{B},
\end{equation}

where

\begin{equation}
\label{eq:3}
d^{(4)} \sigma^{B} = d^{(4)}\Phi^{m}|M_{m}|^{2}F_{j}^{(m)}.  
\end{equation}

Here, $d^{(4)}\Phi^{m}$ and $M_{m}$ respectively denote the phase space element of $m$ particles, taken in four dimensions and the tree level QCD matrix element for the process under consideration. $F_{j}^{(m)}$ is a function of cuts defining the jets etc.. In order to obtain a meaningful result to be compared with
experimental data, typical isolation cuts are applied on the outgoing particles, which may also serve the purpose of keeping the integral finite. A typical criterion for example is to identify outgoing partons with jets and thus apply jet definition cuts on the partons such that they are all well separated in phase space.

The NLO part of the cross section is written as follows

\begin{equation}
\label{eq:4}
\sigma^{NLO} =\int_{m}d^{(d)} \sigma^{NLO}=\int_{m+1}d^{(d)}\sigma^{R}+\int_{m}d^{(d)}\sigma^{V},
\end{equation}

Where $d^{(d)}\sigma^{R}$ denotes the real correction, which consists in emissions of an additional parton and $d^{(d)}\sigma^{V}$ denotes the virtual one-loop corrections. The ultraviolet singularities (UV) in the virtual contribution are treated quite simply: the loop amplitudes are regularized in the dimensional regularization, and the theory is renormalized by the addition of counter-terms in the QCD Lagrangian. The key idea is that we redefine the different physical quantities using these counter-terms in the way become finite observables and to replace the naked parameters with renormalized parameters so that the Lagrangian can be written as the sum of a renormalized part and of a part of counter-terms.

While the two integrals on the right-hand side of the previous equation  are separately inferred divergent in four dimensions, and are therefore taken in d dimensions.  For the real correction, the divergences arise when the additional parton becomes soft or collinear with some other parton, leading to on-shell propagators
in the matrix element. For the virtual correction, the divergence comes with the integration over the unrestricted loop momentum, such that again a propagator goes on-shell. More specifically, infrared safety demands that $F_{j}^{m+1} \rightarrow F_{j}^{m}$ where $m+1-$parton and $m-$parton configurations become kinematically degenerate~\cite{ref:38, ref:39}. 
Setting $d = 4 + 2\epsilon$ in the following, the divergences will manifest themselves in double and single poles, i.e. as $1/\epsilon^{2}$ and $1/\epsilon$, respectively. 

\section{Computational details }
\label{sec:3} 
In this part, we will give some attention to process $ 2 \rightarrow 2 $, $ 2 \rightarrow 3 $ and $ 2 \rightarrow 4$. Thus we choose to study the following process:
$ p p  \rightarrow W^{+} W^{-} $, $ p p  \rightarrow W^{+} W^{-} j $ and $ p p  \rightarrow W^{+} W^{-} j j $ with $j\neq b,t$. We work in the 5-flavour scheme. We provide the cross sections for LHC collisions at center-of-mass energies of $ \sqrt{s} = $14 TeV and an integrated luminosity of $100 fb^{-1}$. 

To implement the computation, we use the following SM parameters

$m_{W}$ = 80.419 GeV, \qquad \qquad $m_{Z}$ = 91.188 GeV,

$m_{H}$ = 125 GeV, \qquad \qquad $G_{\mu}$ = 1.1664 $\times$ $10^{-5}$ GeV.

We fix  central  values of the renormalization and factorization scales to the Z mass $\mu_{R}$ = $\mu_{F}$ = $m_{Z}$ and the scale uncertainties are estimated in the interval 0.5 $m_{Z} \leq \mu_{R}$, $\mu_{F} \leq $ 2 $m_{Z}$ (our choice is by default).

The results are evaluated both at the LO and NLO accuracy in QCD using the NNPDF23 parton distribution functions (PDFs). These pdfs systematically include all relevant data from the LHC by dint of the excellent performance of the LHC and its experiments. In NNPDF23 framework, the values of the strong coupling $ \alpha_{s}(M_{Z}) $ at LO and NLO are 0.130 and 0.118 respectively.

We use the MC@NLO method~\cite{ref:40} as implemented in MadGraph5$\_$aMC@NLO, in order to produce events first, at the partonic level before moving to a study based on  (PS) parton shower and including hadronisation. 
To this aim, the realistic descriptions of LHC collisions require to match hard scattering matrix elements to a modeling of QCD environment, we choose to match the hard scattering processes with Pythia8~\cite{ref:41}  parton showering and hadronization as part of the Monte Carlo simulation, which is one of the string model. The obtained events are run through the fast detector simulation, Delphes~\cite{ref:42}, which serves to accurately model the ATLAS and CMS detectors and the final state reconstructions that are performed. The reconstruction is done using the ATLAS$\_$1604$\_$07773.tcl card which is based on recent experimental analyzes of the LHC. Jet reconstruction is then done by means of the anti-$k_T$ algorithm~\cite{ref:43} with a radius parameter set to R = 0.6 as included in the FastJet program~\cite{ref:44} and we impose two cuts on jet transverse momentum,

\begin{equation}
\label{eq:5}
p_{ T}(j) > 20  GeV,  \qquad    p_{ T}(j) > 100   GeV
\end{equation}

\begin{equation}
\label{eq:6}
\ |\eta(j)| \leq 2.5 \qquad and  \qquad \Delta R_{j_{1}j_{2}}= \sqrt{(\eta_{j_{1}} -\eta_{j_{2}})^{2}  + (\phi_{j_{1}} - \phi_{j_{2}})^{2}} =0.5
\end{equation}
with $\eta(j)$ and $\phi_{j}$  are respectively the pseudo-rapidity and the azimuth of particle j.

These cuts are typical of those used at the LHC. The events are finally analysed with the MadAnalysis5 package ~\cite{ref:45}.

\subsection{LO total cross section}
\label{subsec:1}

\begin{figure}[]
	\centering	
	\includegraphics[width=.30\textwidth,trim=0 0 0 0,clip]{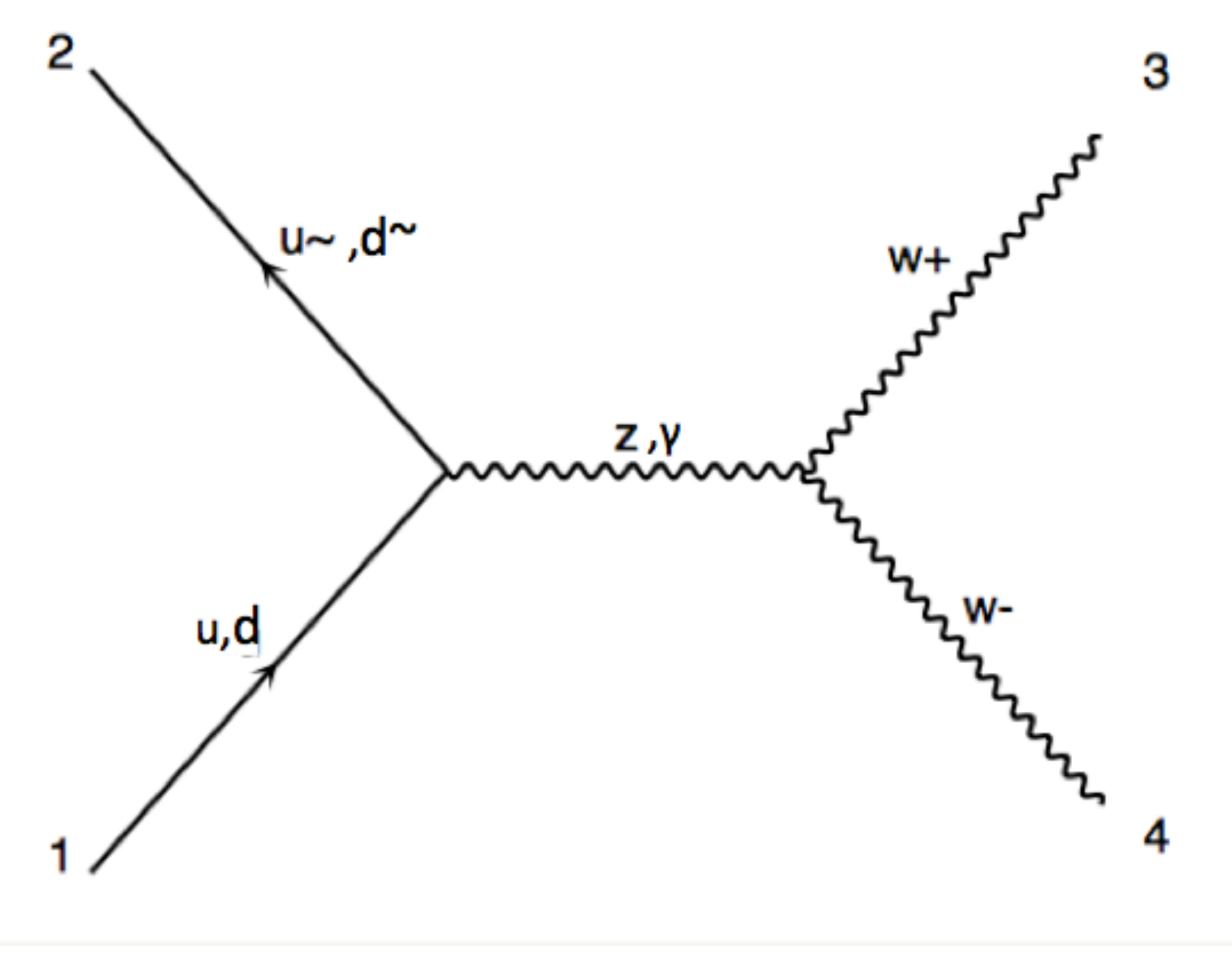}
		\includegraphics[width=.30\textwidth,trim=0 0 0 0,clip]{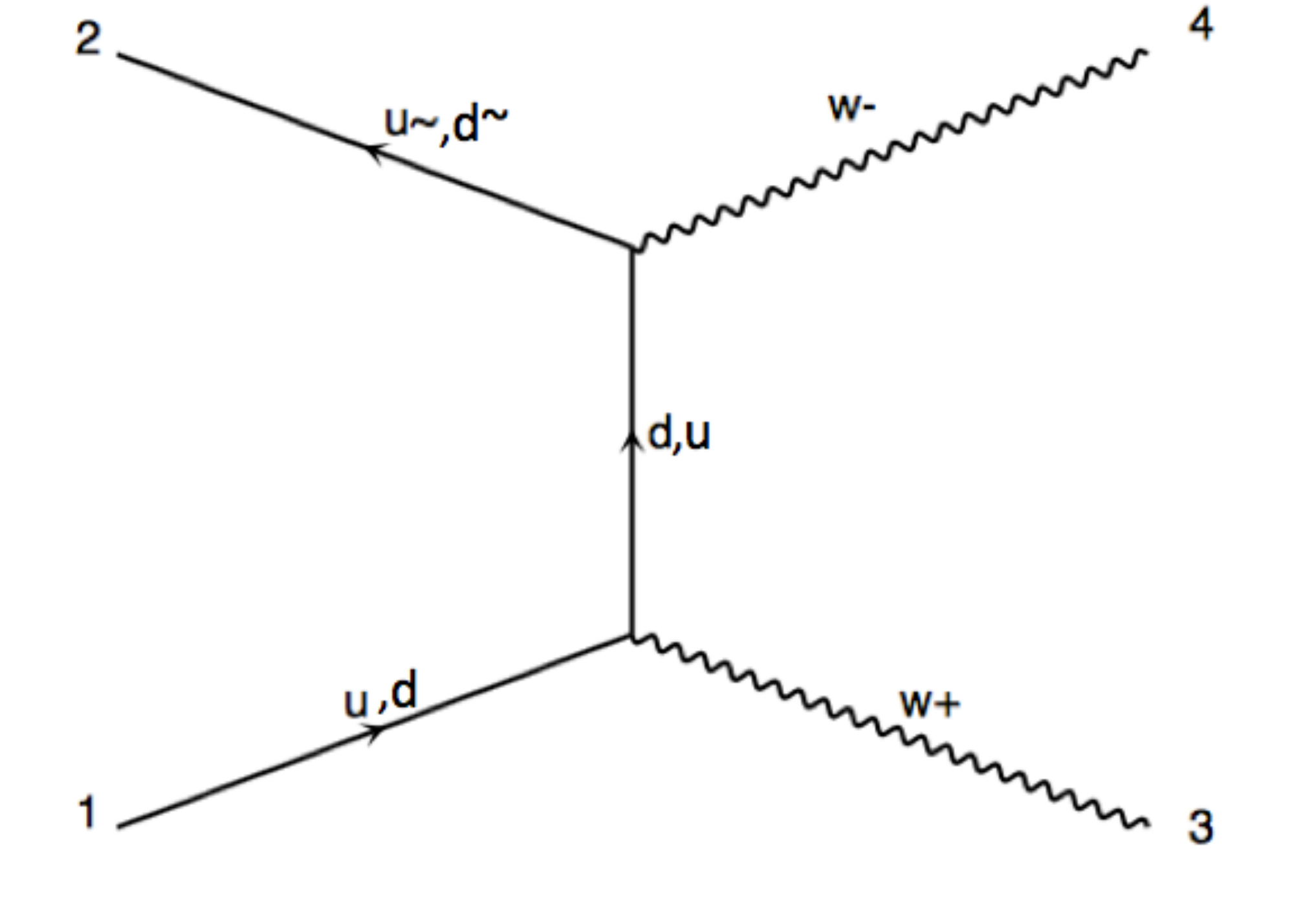} \\	
	\qquad 	a- $ p p  \rightarrow W^{+} W^{-} $ 
	
	\includegraphics[width=.30\textwidth,trim=0 0 0 0,clip]{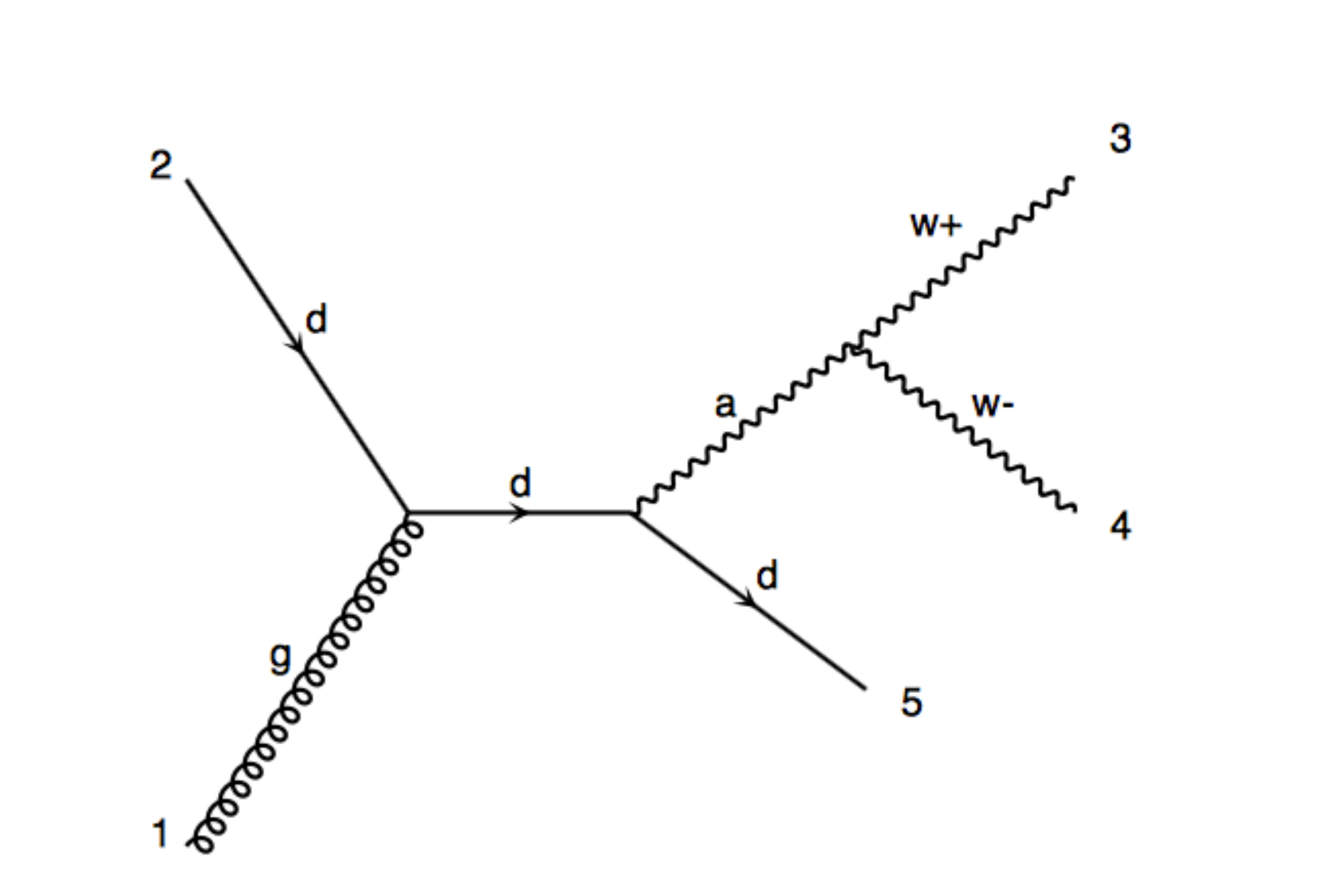} 	\includegraphics[width=.30\textwidth,trim=0 0 0 0,clip]{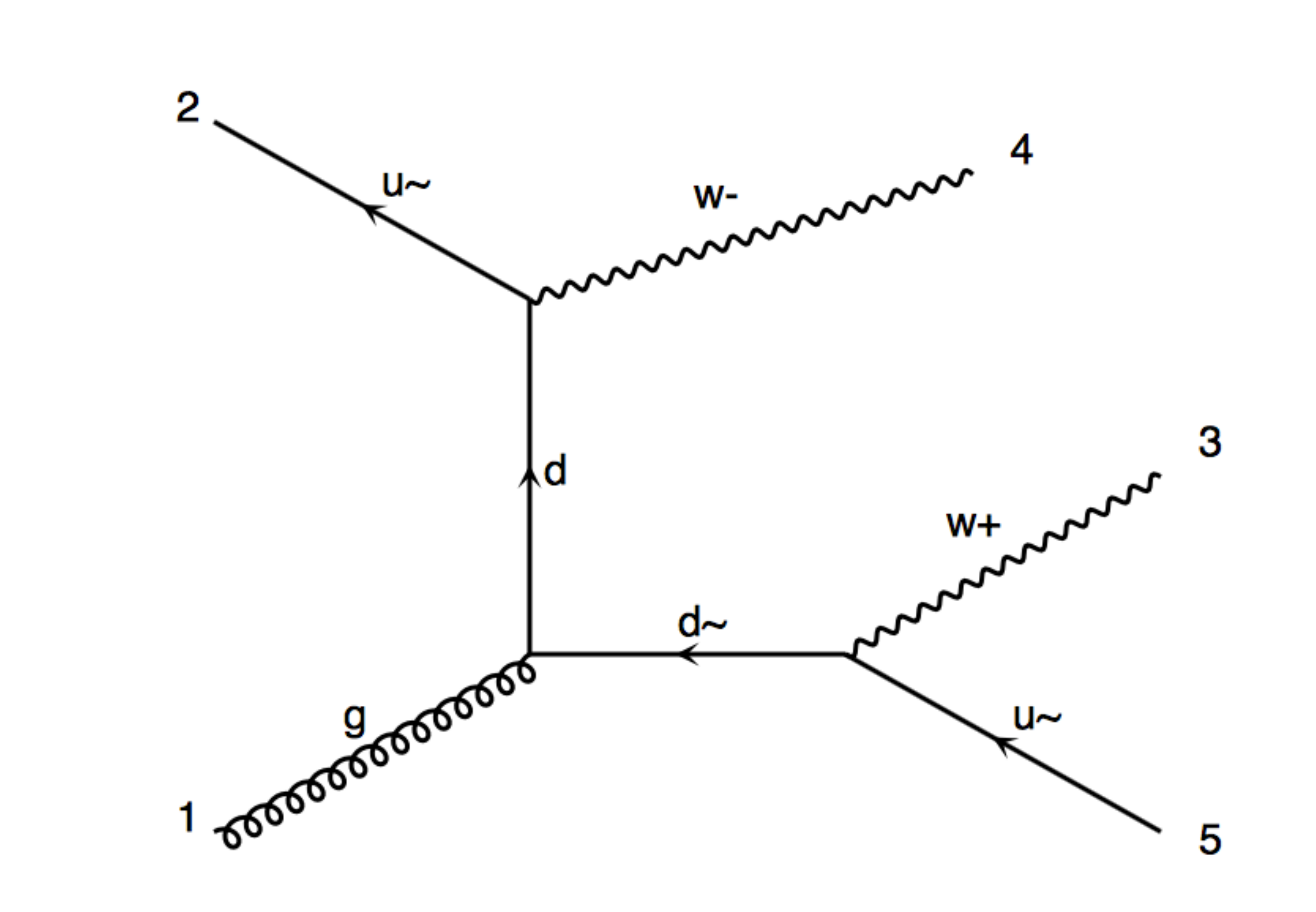} 	\includegraphics[width=.30\textwidth,trim=0 0 0 0,clip]{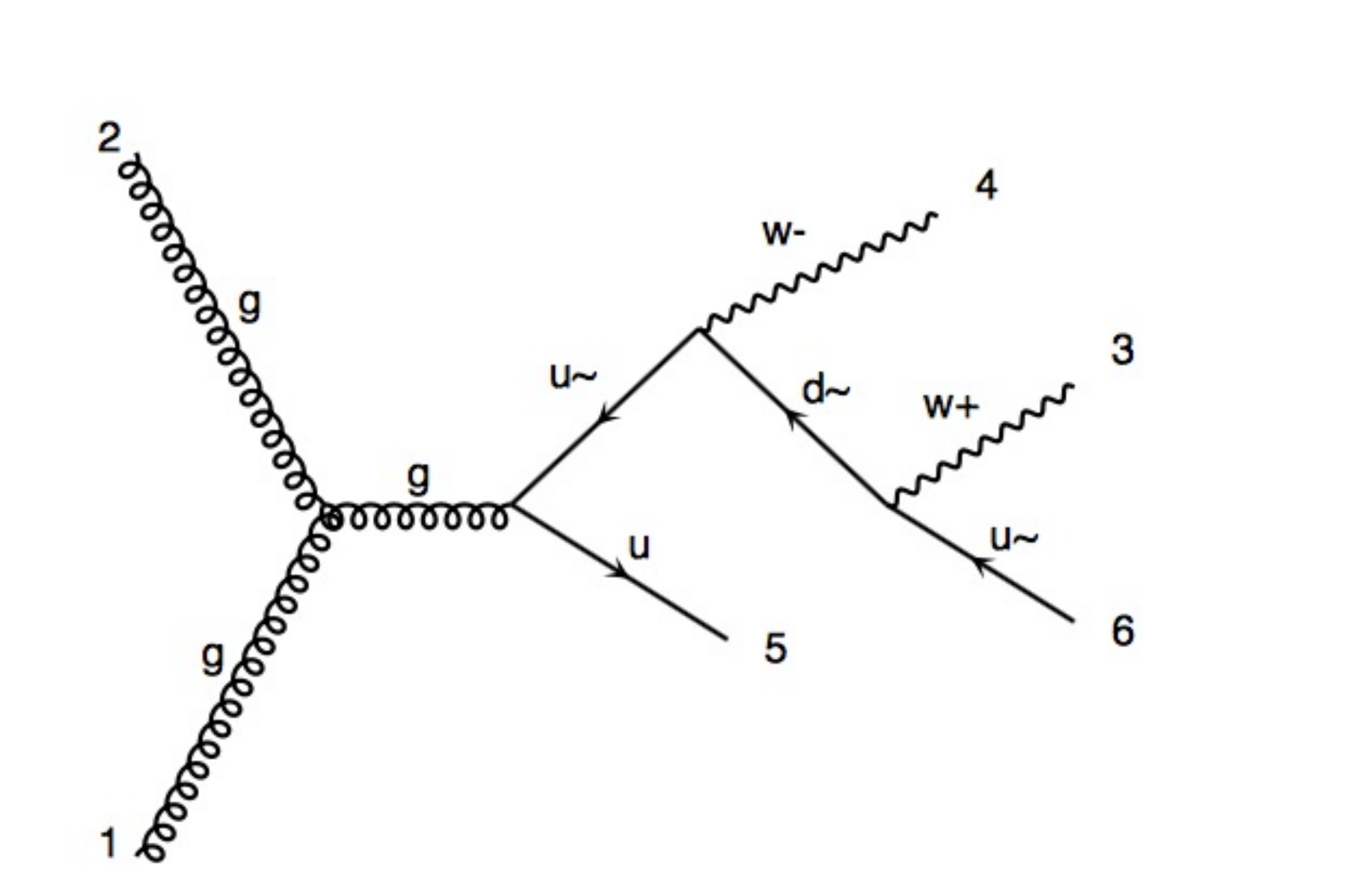}\\
	\qquad 	\qquad 	\qquad 	b- $ p p  \rightarrow W^{+} W^{-}j $ 	\qquad 	\qquad 	\qquad  	\qquad  	c- $ p p  \rightarrow W^{+} W^{-}j j$ 
	\caption{Examples of Feynman diagrams for the different processes of  production of a pair of $W^{+} W^{-} $ bosons  in association with 0, 1 and 2 jets at tree-level.\label{fig:1}
	}	
\end{figure}

We start our study by generating our processes at leading order(LO). At this level, the dominant contribution for the process $ p p  \rightarrow W^{+} W^{-}$  comes from the quark-antiquark ($q\bar{q}$) annihilations via the $t$-channel with a Z or (and) $\gamma$ exchange  and the $s$-channel  with a quark exchange as shown in figure~\ref{fig:1}.a.
While the production of $W^{+}W^{-}j$ is generated by $q \bar q$, $qg$ and $\bar q g $ , an example of such contributions is illustrated in figure~\ref{fig:1}.b. The $pp\longrightarrow W^{+}W^{-}jj$ gets a further contribution from $gg$ as presented in figure~\ref{fig:1}.c. At the tree-level, the total cross sections are given by the equation \ref{eq:2}.

\begin{table}[]
	\centering
	\begin{tabular}{|c|c|c|}
		\hline
		
		&	\multicolumn{1}{|c|} {$p^{min}_{T,j} > 20$ GeV}  & \multicolumn{1}{|c|}{$p^{min}_{T,j} > 100$ GeV}  \\ 	\hline 
		
		&	$ \sigma_{LO}$[pb]   & $ \sigma_{LO}$[pb]     \\  \hline

		$W^{+}W^{-}j$   &  40.54 $ \pm$0.086 $_{-9.8\%}^{+11.1\%}$  &  7.651$ \pm$0.025 $_{-10.3\%}^{+12\%}$  \\ \hline 
		
		$W^{+}W^{-}jj$      & 21.51 $ \pm$0.042 $_{-17.6\%}^{+23.6\%}$ &   1.847  $ \pm$0.004 $_{-19.2\%}^{+25.7\%}$  \\ \hline	
		
	\end{tabular} 
	\caption{The LO  total cross section for the various  processes of production at  $\sqrt{s}= 14$ TeV, at partonic level for different cuts on jet transverse momentum $p_{T,j} > 20$ GeV and $p_{T,j} > 100$ GeV. }
	\label{tab:1}
\end{table}

The  LO total cross section of $W^{+} W^{-} $ pairs production without jet is $\sigma_{p p  \rightarrow W^{+}W^{-}}^{LO} = $ 66.07 $\pm $ 0.03(stat) $_{-6.8\%}^{+5.8\%}$ (syst) pb. Our predictions are close to that published in the in the ref~\cite{ref:34}, where the authors have used MSTWnlo2008 PDFs and obtained $\sigma_{ LO}$ = 73.55 $\pm $ 0.005 $_{-6.1\%}^{+5.0\%}$ pb at $\sqrt{s}$ = 13 TeV. 

For $ p p  \rightarrow W^{+} W^{-} j$ and  $p p  \rightarrow W^{+} W^{-} jj$, the accompanying jets are subject to cuts. We consider two cuts on jets transverse momentums, $p_{T,j} > 20$ GeV and $p_{T,j} > 100$ GeV. The corresponding cross-sections are presented in table~\ref{tab:1}, we observe that these cuts have a significant influence on the measurements of LO total cross sections of $W^{+} W^{-} $ pairs production with 1 and 2 jets. The numerical predictions of table~\ref{tab:1}  show that the total cross section at $p^{min}_{T,j} > 20$ GeV is roughly a factor of 5 to 10 larger than those at  $p^{min}_{T,j} > 100$ GeV. These predictions are approximately in agreement with the one found in ref~\cite{ref:18} where they used the CTEQ6 PDFs at $ \sqrt{s}$ = 14 TeV and ref~\cite{ref:34} uses the MSTWnlo2008 PDFs  at $ \sqrt{s} $ = 13 TeV with a cut on $p_{T,j} > 30$ GeV, different from ours that's why we have more events what gives a slightly larger total cross section.

\subsection{ NLO total cross sections}
\label{subsec:2}

\begin{figure}[]
	\centering	
	
	\includegraphics[width=.30\textwidth,trim=0 0 0 0,clip]{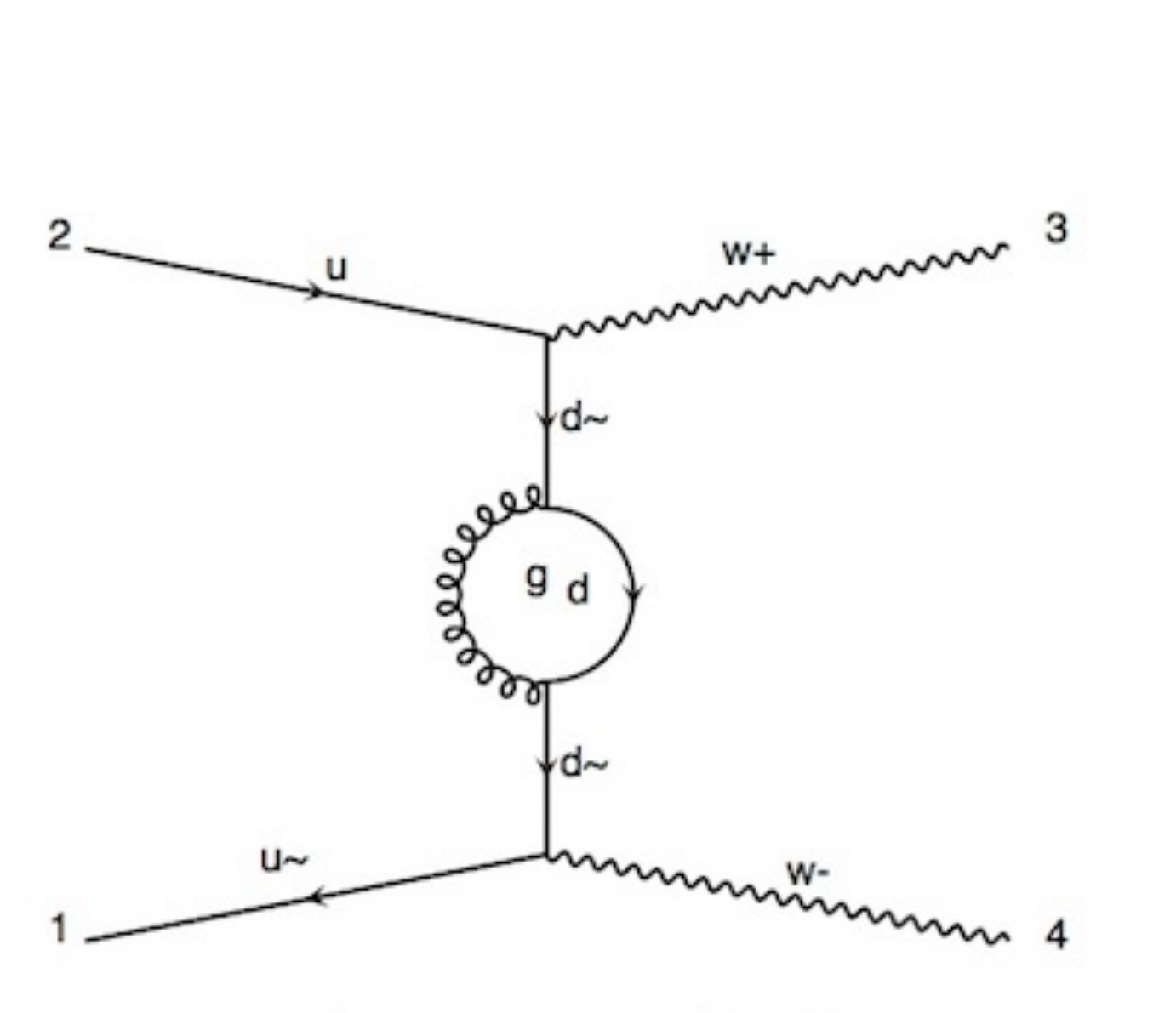} 	\includegraphics[width=.30\textwidth,trim=0 0 0 0,clip]{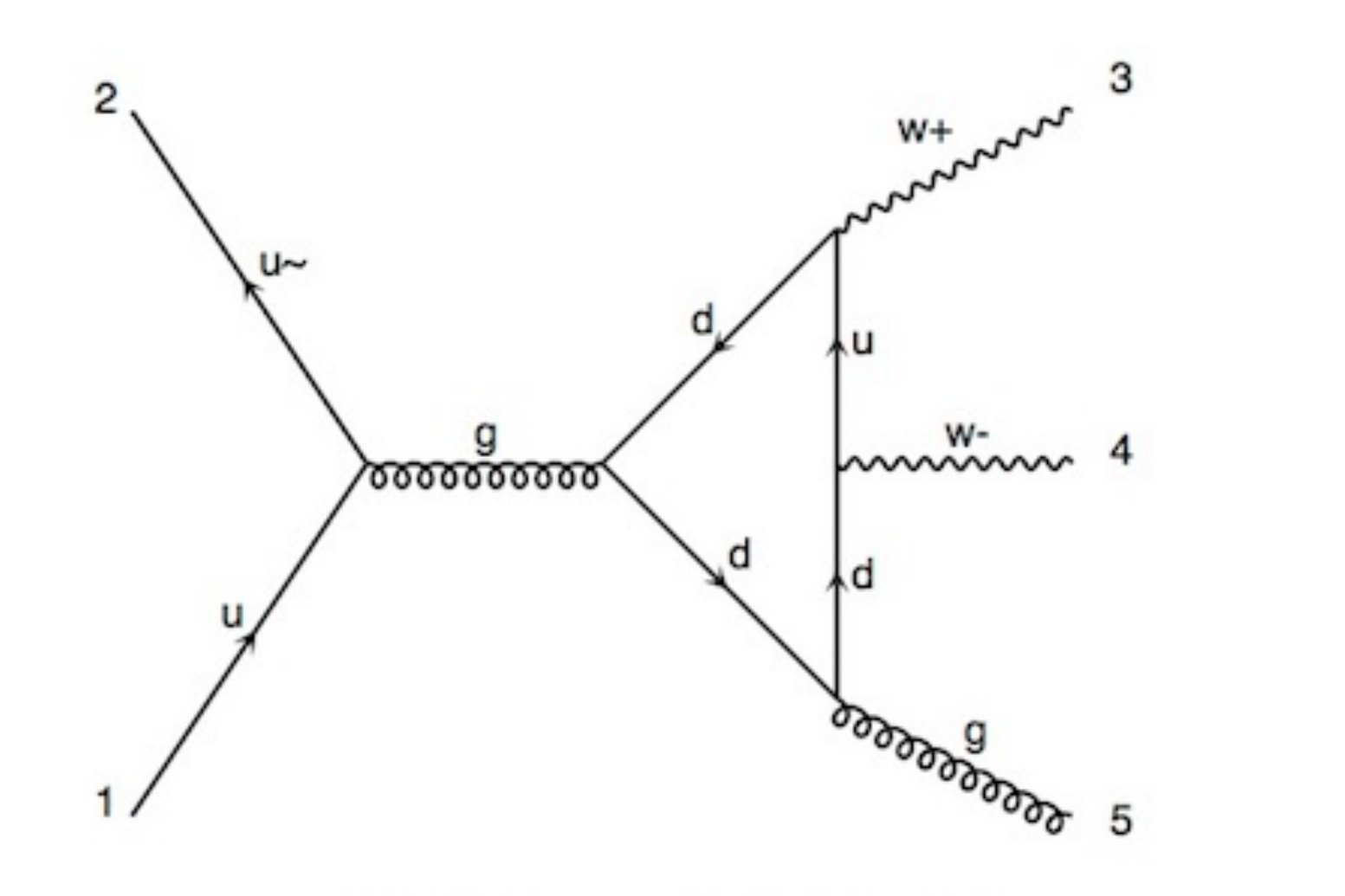}
	\includegraphics[width=.30\textwidth,trim=0 0 0 0,clip]{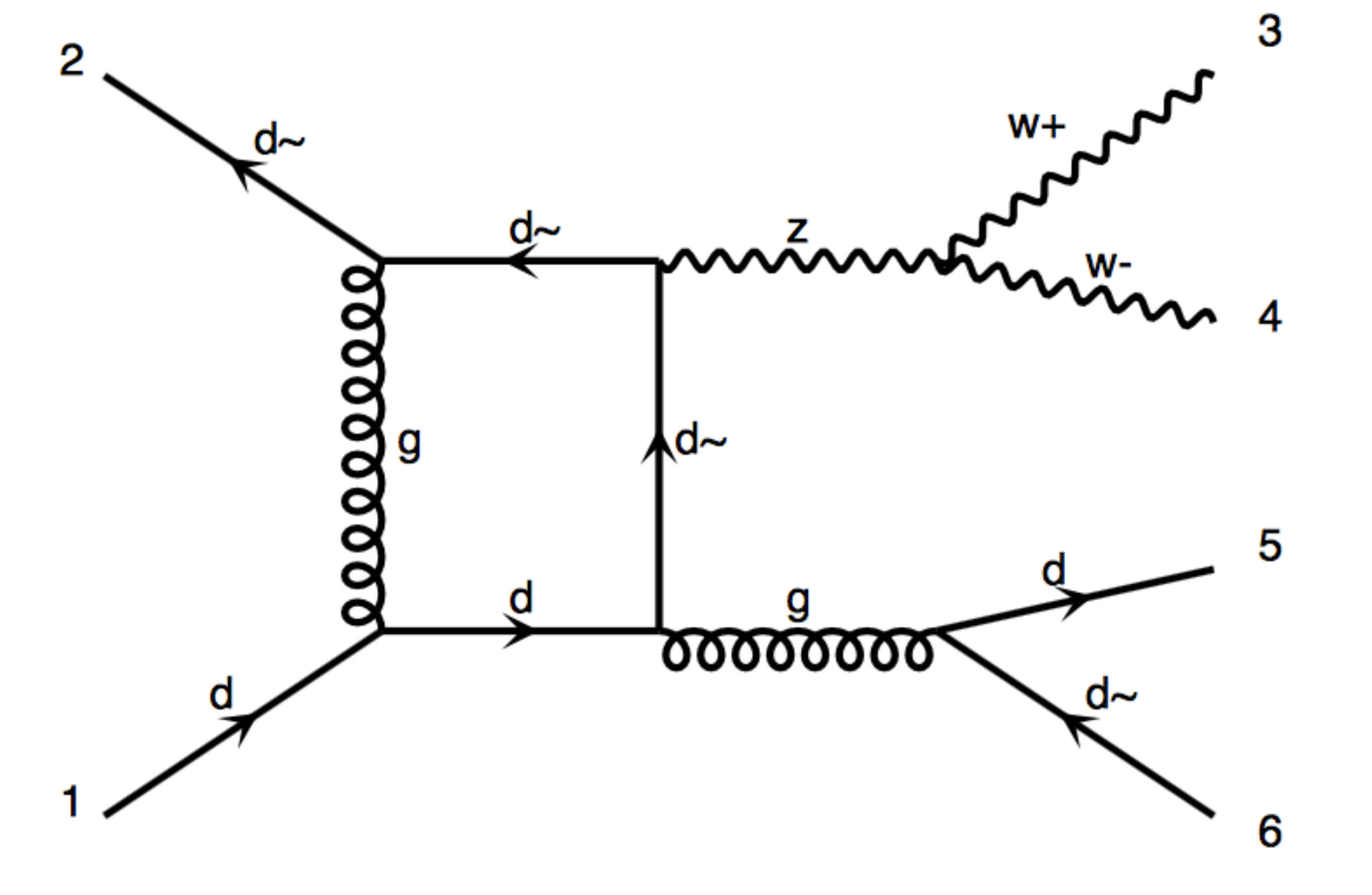} \\
	a- $ p p   \rightarrow W^{+} W^{-} $    \qquad \qquad	b- $ p p   \rightarrow W^{+} W^{-} j$   \qquad \qquad c- $ p p  \rightarrow W^{+} W^{-} jj$  
	\caption{ Examples of Feynman diagrams of virtual corrections for the different processes of  production of a pair of $W^{+} W^{-} $ bosons  in association with 0, 1 and 2 jets. 
	}\label{fig:2}	
\end{figure}

\begin{figure}[]
	\centering	
	
	\includegraphics[width=.30\textwidth,trim=0 0 0 0,clip]{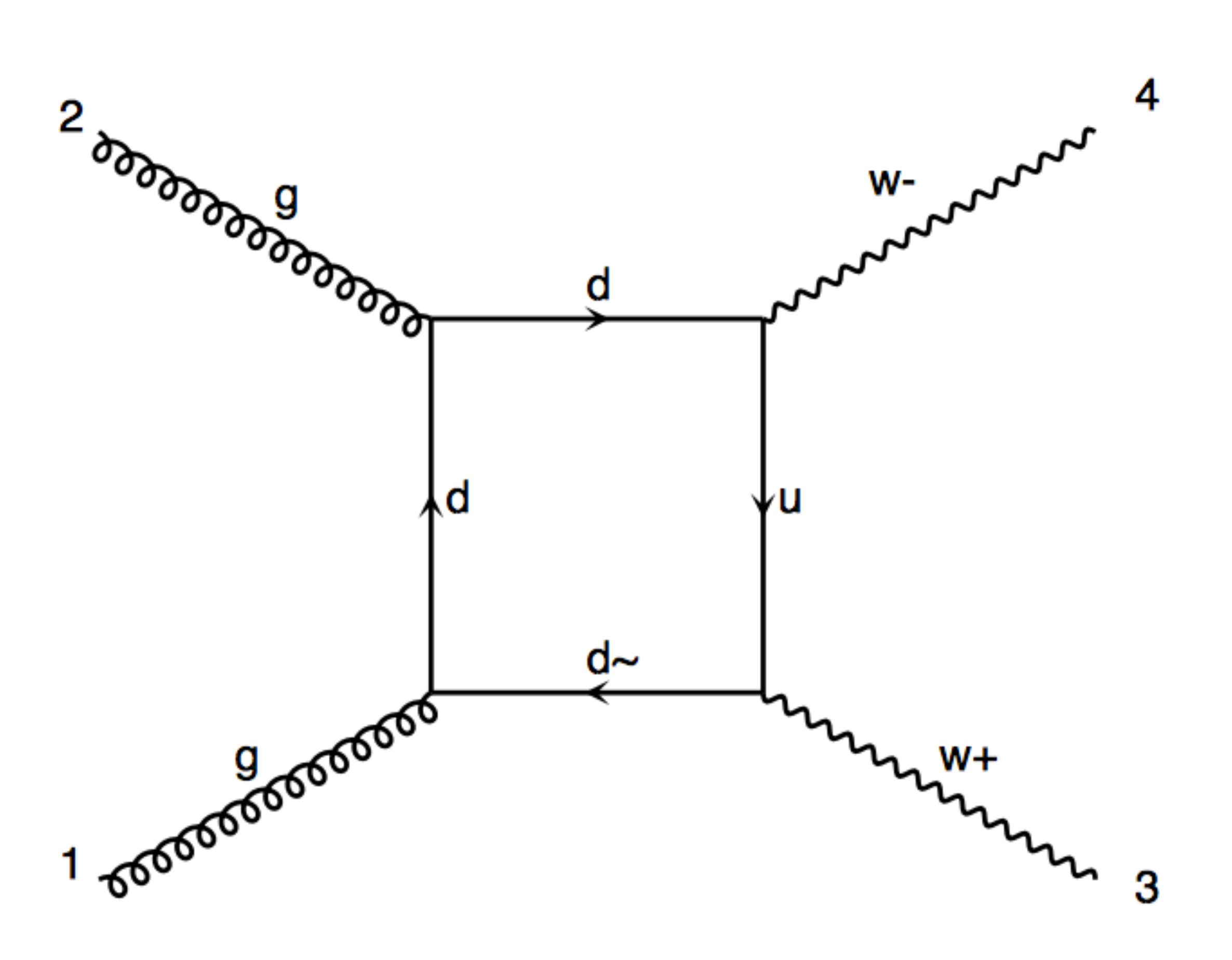} 	\includegraphics[width=.30\textwidth,trim=0 0 0 0,clip]{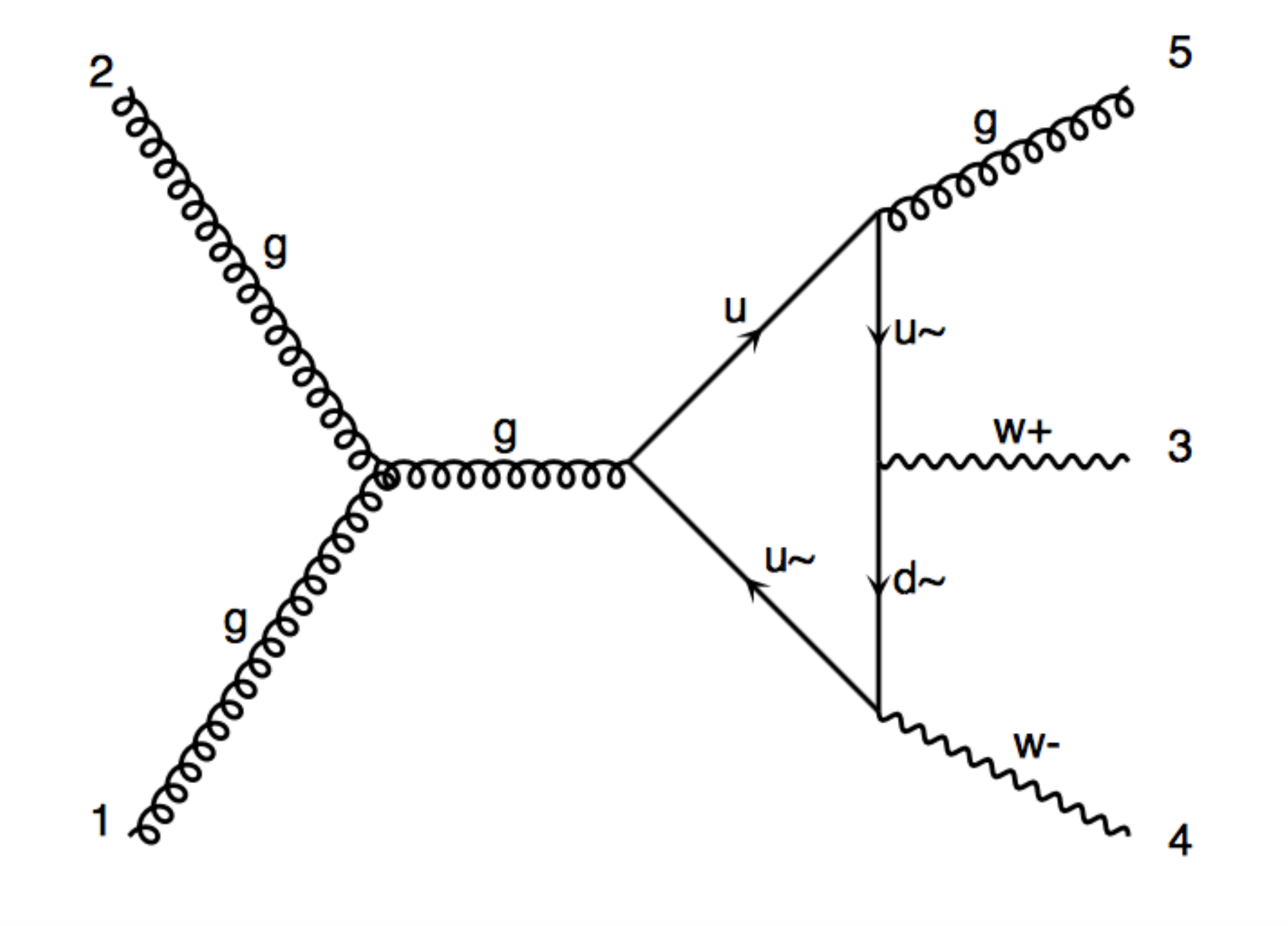}
	\includegraphics[width=.30\textwidth,trim=0 0 0 0,clip]{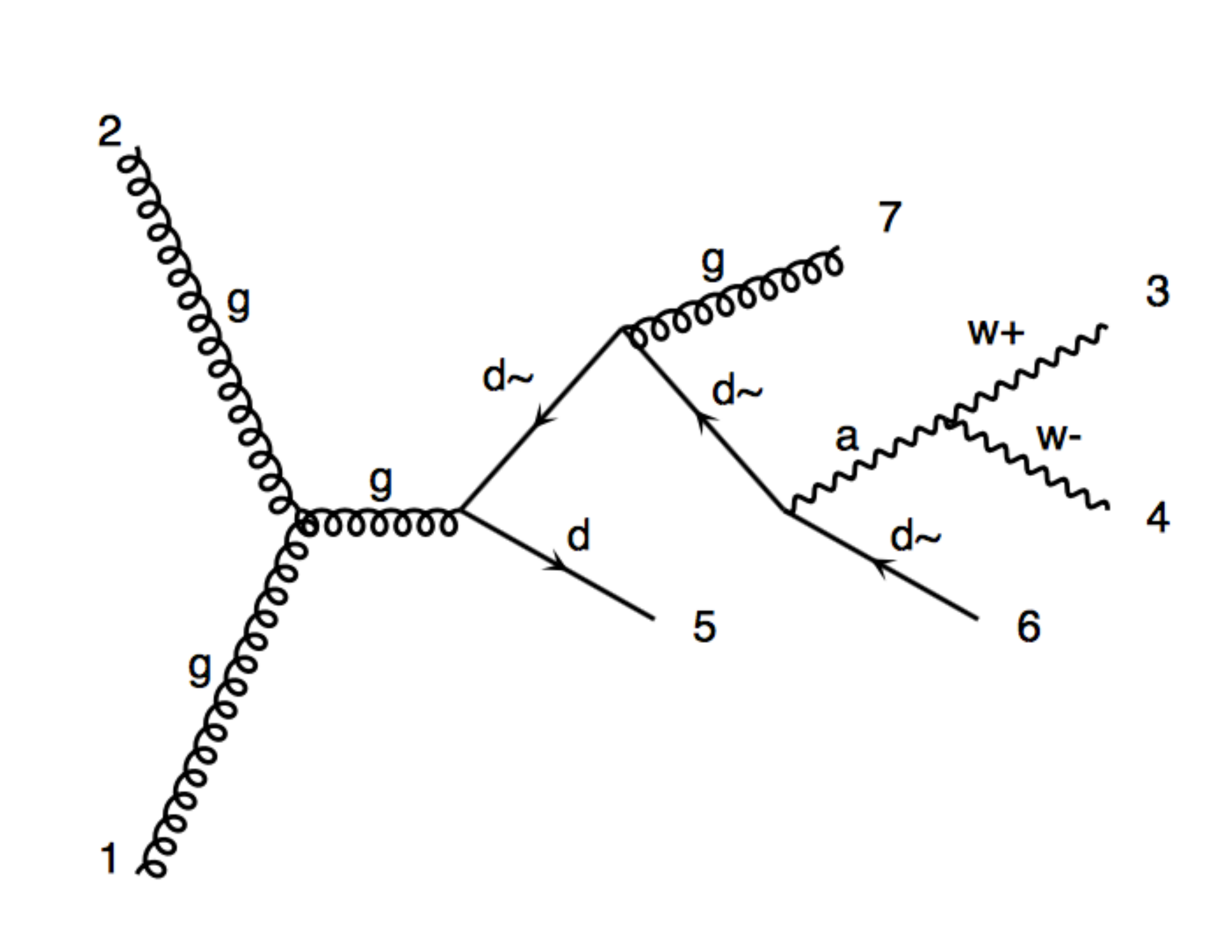} \\
	a- $ p p   \rightarrow W^{+} W^{-} $    \qquad \qquad	b- $ p p   \rightarrow W^{+} W^{-} j$   \qquad \qquad c- $ p p  \rightarrow W^{+} W^{-} jj$  
	\caption{ Examples of Feynman diagrams of gluons fusion for the different processes of  production of a pair of $W^{+} W^{-} $ bosons  in association with 0, 1 and 2 jets. 
	}\label{fig:3}	
\end{figure}

At the next leading order, there are more diagrams than at the tree-level ones for all our processes. This is on account of the real ( Born contribution with additional parton in the final state) and the virtual contributions as shown in figure~\ref{fig:2}. In addition gg fusion diagrams contribute at this level for the processes as shown in figure~\ref{fig:3}

For the process $ p p  \rightarrow W^{+} W^{-}$, the NLO total cross section obtained is: $ \sigma_{p p  \rightarrow W^{+}W^{-}}^{NLO}$ = 115.114 $\pm $ 0.4 (stat) $_{-4.4\%}^{+3.9\%}$ (syst) pb, while the gluon fusion contribution is $ \sigma_{g g \rightarrow W^{+}W^{-}}^{NLO}$ = 3.604$\pm $  0.0061(stat) $_{-16.6\%}^{+20.7\%}$ (syst) pb, it represents only 3.13$\% $ of the total cross section. It is important to stress that $WW$ at NLO includes also a jet as a result of the real emission. This jet is not subject to any cut. When studying $WWj$ at NLO, real emission includes an extra jet which we differentiate from the LO $WWj$ on which cuts are imposed.

\begin{table}[]
	\centering
	\begin{tabular}{|c|c|c|c|}
		\hline
		
		&	\multicolumn{3}{|c|}{$p^{min}_{T,j} > 20$ GeV} 	 \\ 	\hline 	
		&	$\sigma_{NLO}$[pb]   &$  K$ & $\sigma_{gg}$[pb]    \\  \hline 
		
		$W^{+}W^{-}j$   &     59.83 $ \pm$0.27  $_{-5.0\%}^{+4.5\%}$  & 1.48  & 1.67 $ \pm$ 0.0029 $_{-25.4\%}^{+37.6\%}$  \\ \hline
		$W^{+}W^{-}jj$      &  27.77 $ \pm$0.19 $_{-5.0\%}^{+1.9\%}$&1.28 & 1.099 $ \pm$ 0.0092  $_{-21.9\%}^{+30.2\%}$   \\ \hline	
		
	\end{tabular} 
	\caption{The  NLO total cross section for the various  processes of production at  $\sqrt{s}= 14$ TeV, at partonic level with cuts on jet transverse momentum $p_{T,j} > 20$ GeV. }
	\label{tab:2}
	
	\centering
	\begin{tabular}{|c|c|c|c|}
		\hline
		
		&	\multicolumn{3}{|c|}{$p^{min}_{T,j} > 100$ GeV} 	 \\ 	\hline 	
		&	$\sigma_{NLO}$[pb]   &$  K$ & $\sigma_{gg}$[pb]    \\  \hline 
		
		$W^{+}W^{-}j$   &    12.123 $ \pm$0.047 $_{-5.6\%}^{+6.0\%}$	 & 1.58 & 0.0885 $ \pm$ 0.00013 $_{-27.4\%}^{+41.4\%}$ \\ \hline 
		$W^{+}W^{-}jj$      & 2.497 $ \pm$0.015 $_{-7.5\%}^{+4.4\%}$&  1.35 & 0.071 $ \pm$ 0.0004$_{-23.9\%}^{+34.0\%}$\\ \hline	
	\end{tabular} 
	\caption{The  NLO total cross section for the various  processes of production at  $\sqrt{s}= 14$ TeV, at partonic level with cuts on jet transverse momentum $p_{T,j} > 100$ GeV. }
	\label{tab:3}
	
\end{table}

Numerical values of total cross section predictions for the $W^{+}W^{-}$ pairs production with one and two jets are represented in the table \ref{tab:2} and the table \ref{tab:3}. Similarly to LO predictions,  the two cuts considered have an important effect on the measurements of NLO total cross section, where we clearly see that the NLO total cross section at$p_{T,j} > 20$ GeV is greater than at $p_{T,j} > 100$ GeV. The contribution of the gluon fusion is for 2 $ \% $ to 3 $ \% $ of the total cross section for each process. For the $ p p   \rightarrow W^{+} W^{-} $, 76$\%$ of its total cross section is given by $q\bar{q}$ contributions, unlike to the process associated with jets, the largest part of its total cross section(56$\%$ for $ p p   \rightarrow W^{+} W^{-} j$ and 68$\%$ for $ p p   \rightarrow W^{+} W^{-} jj $) is given by the contributions $qg$ and $\bar{q}g$.

We calculate the K factor which is the ratio of the NLO total cross section on LO total cross section($K = \frac{\sigma_{NLO}}{\sigma_{LO}}$). The K factor for the process $ p p   \rightarrow W^{+} W^{-} $ is equal to 1.74. As we can see, the NLO total cross section for all the  processes are greater from 28$\%$ to 74$\%$ than the first order terms. The process $W^{+}W^{-}$ has the largest value, as we can also see that adding the jets to the final state decreases this factor. On the other hand, the table gives an important information that the K factor does not depend only on the choice of the scales (factorization, renormalization), the PDF but also on the cut to perform on the transverse moments. We see that at $p_{T,j} > 100$ GeV, the K factor is more significant that at $p_{T,j} > 20$ GeV.

\section{Numerical results}

\label{sec:4}

\subsection{Partonic level}
\label{subsec:3}

\begin{figure}[]
	\centering	
	
	\includegraphics[width=.45\textwidth,trim=0 0 0 0,clip]{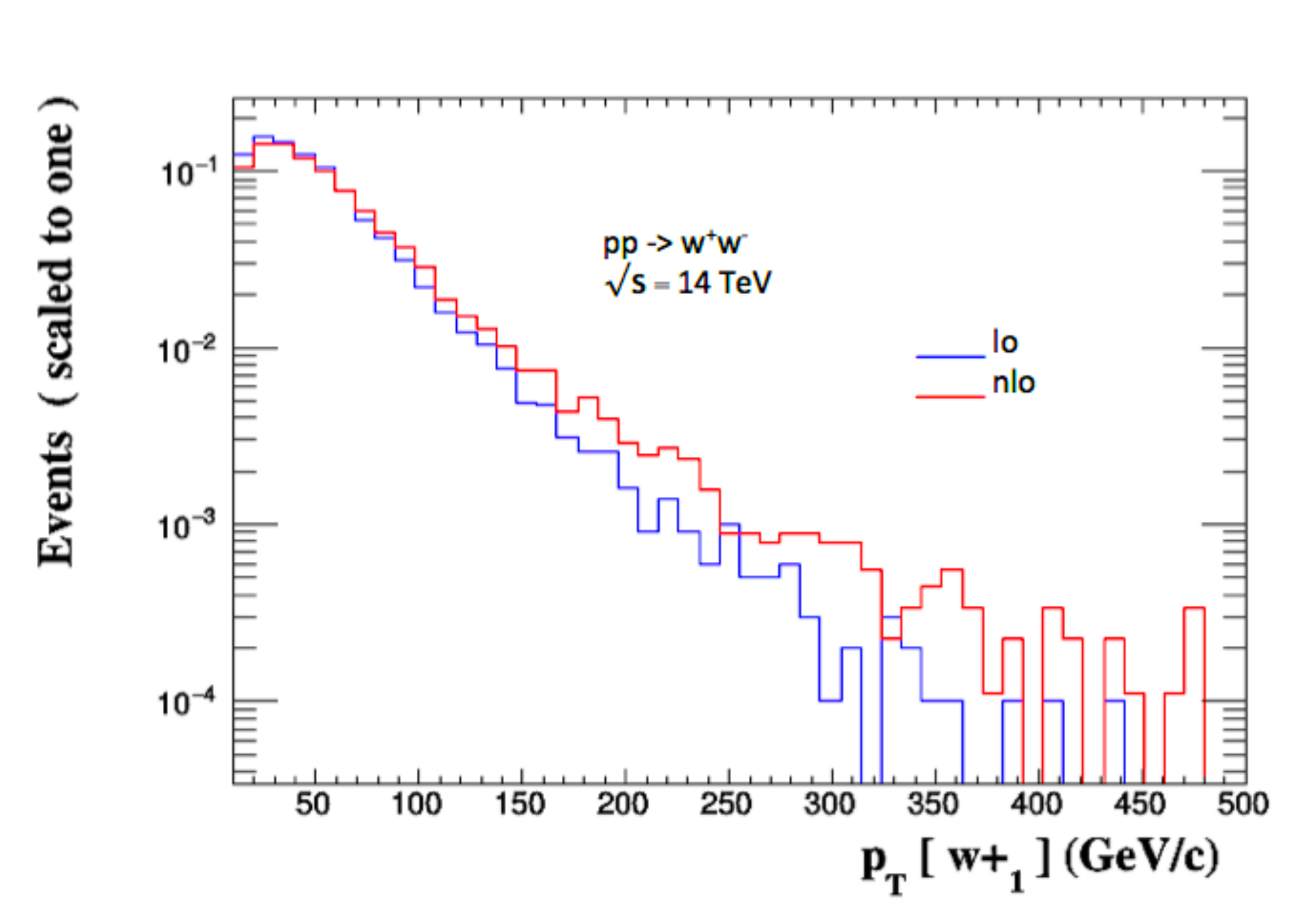} 	
	\includegraphics[width=.45\textwidth,trim=0 0 0 0,clip]{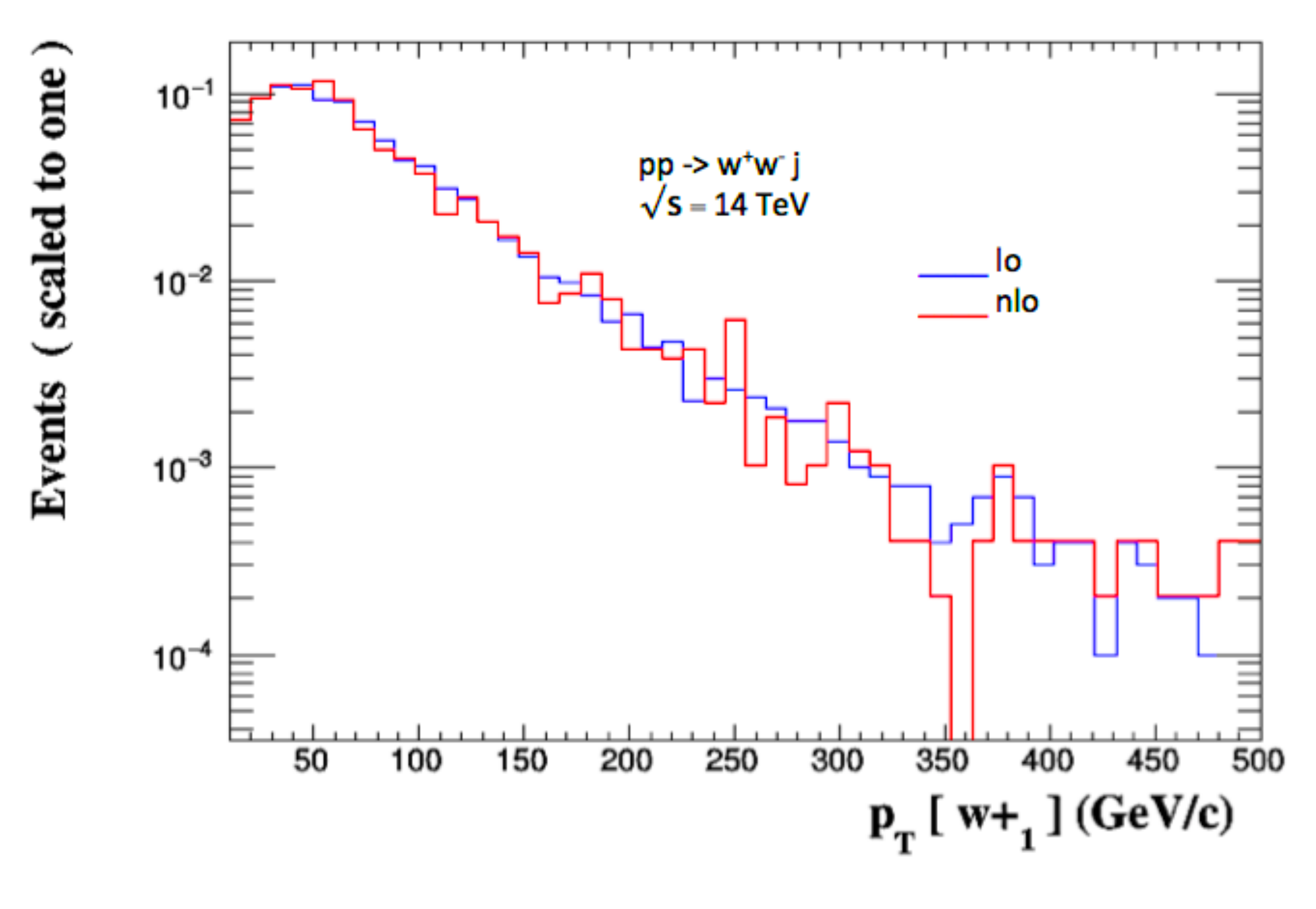} 
	\includegraphics[width=.45\textwidth,trim=0 0 0 0,clip]{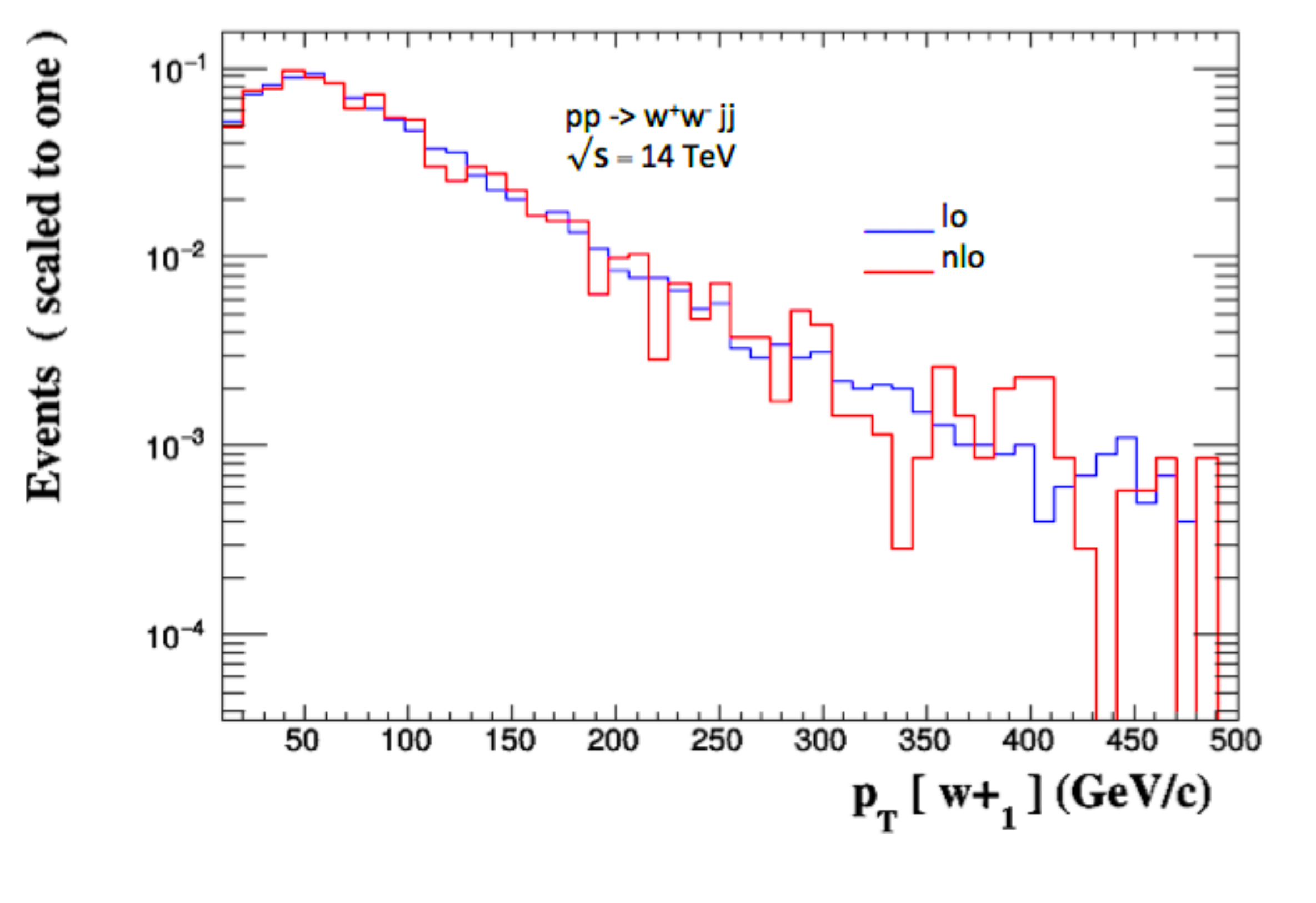} 
	
	\caption{ $W^{+} $ transverse momentum distribution at partonic level for $W^{+}W^{-}$ production with 0, 1 and 2 jets at $\sqrt{s}= 14$ TeV.
	}\label{fig:4}

\end{figure}

We start the discussion at partonic level, with a presentation of the transverse momentum distributions of the harder $W^{+}$ boson $p_{T}(W^{+}_{1})$ at LO and NLO with QCD corrections as show in figure~\ref{fig:4}. In general, the distributions for all our processes, have the same shape where two different regions are distinguished. In the first one, the distribution increases until the maximum at  $p_{T} \approx 60 $ GeV and in the second one at 60 GeV  $ < p_{T} < 500$ GeV, there is a decrease on this observable. Obviously, the same behaviour has been established with the transverse momentum distribution of the harder $W^{-}$ $p_{T}(W^{-})$. 

\begin{figure}[]
\centering	
	
	\includegraphics[width=.45\textwidth,trim=0 0 0 0,clip]{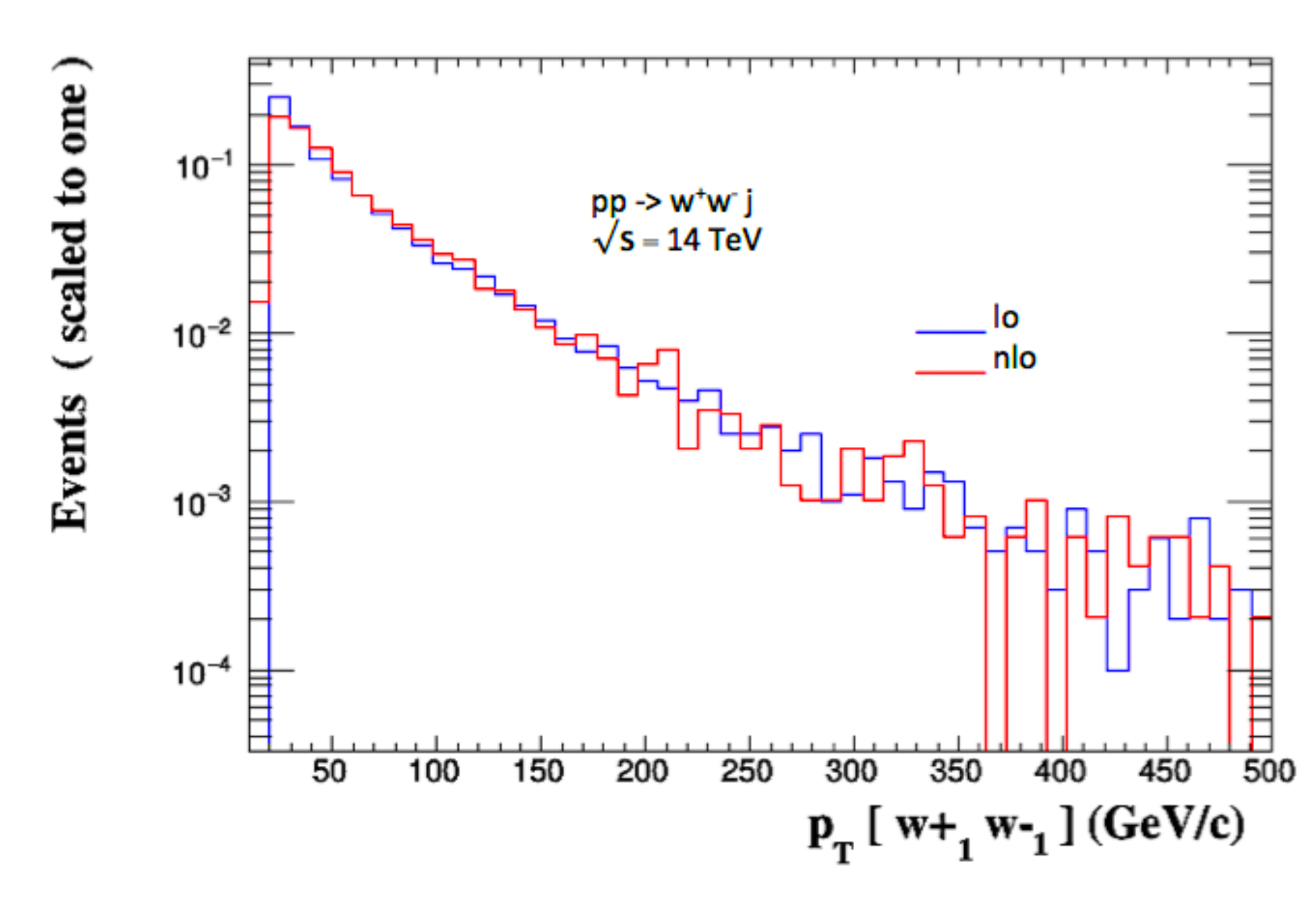} 
	\includegraphics[width=.45\textwidth,trim=0 0 0 0,clip]{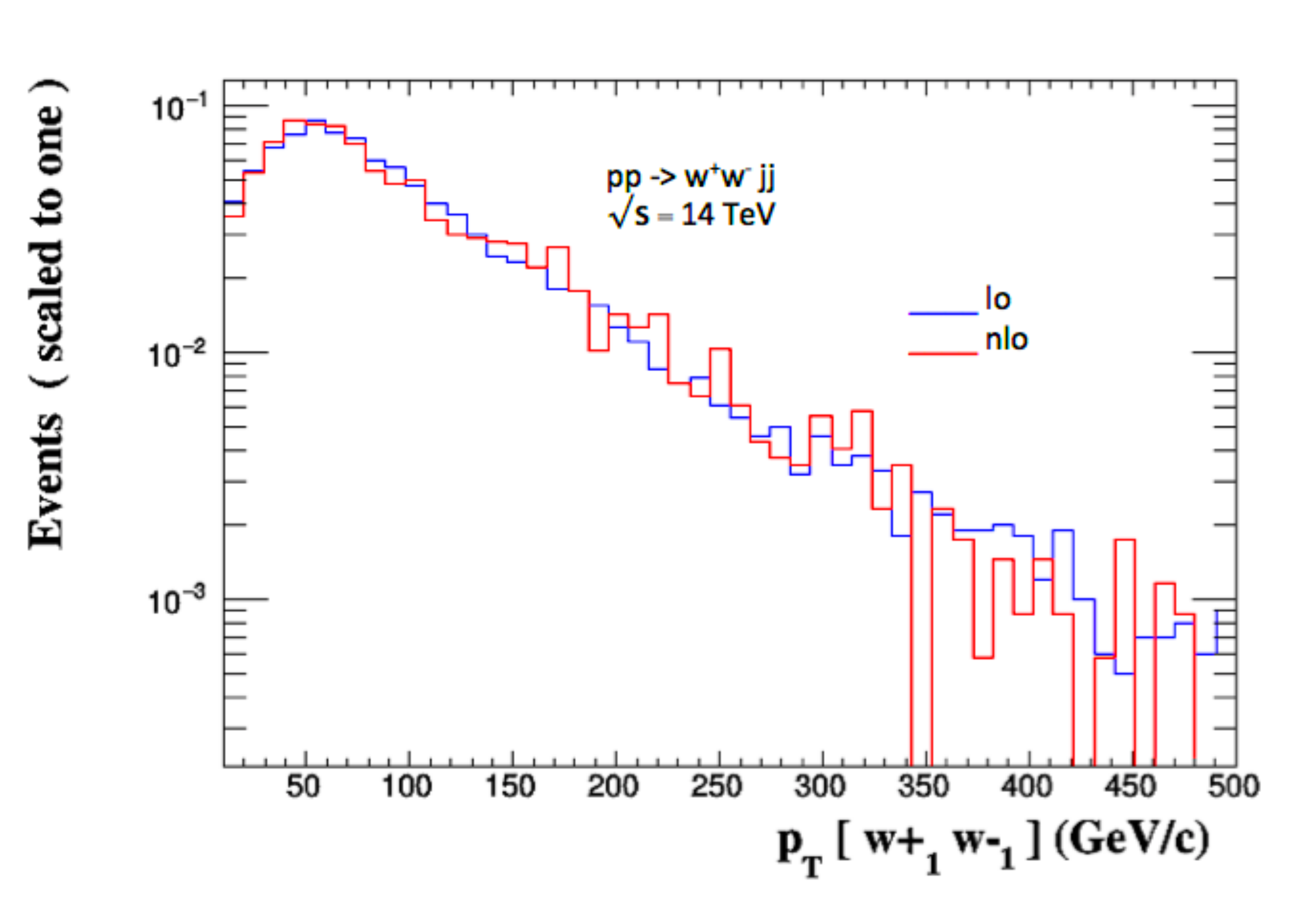} 
	
	\caption{ $W^{+}W^{-} $ transverse momentum distribution at partonic level for $W^{+}W^{-}$ production with 1 and 2 jets at $\sqrt{s}= 14$ TeV.
	}\label{fig:5}

\end{figure}

\begin{figure}[]
\centering

	\centering	
	\includegraphics[width=.45\textwidth,trim=0 0 0 0,clip]{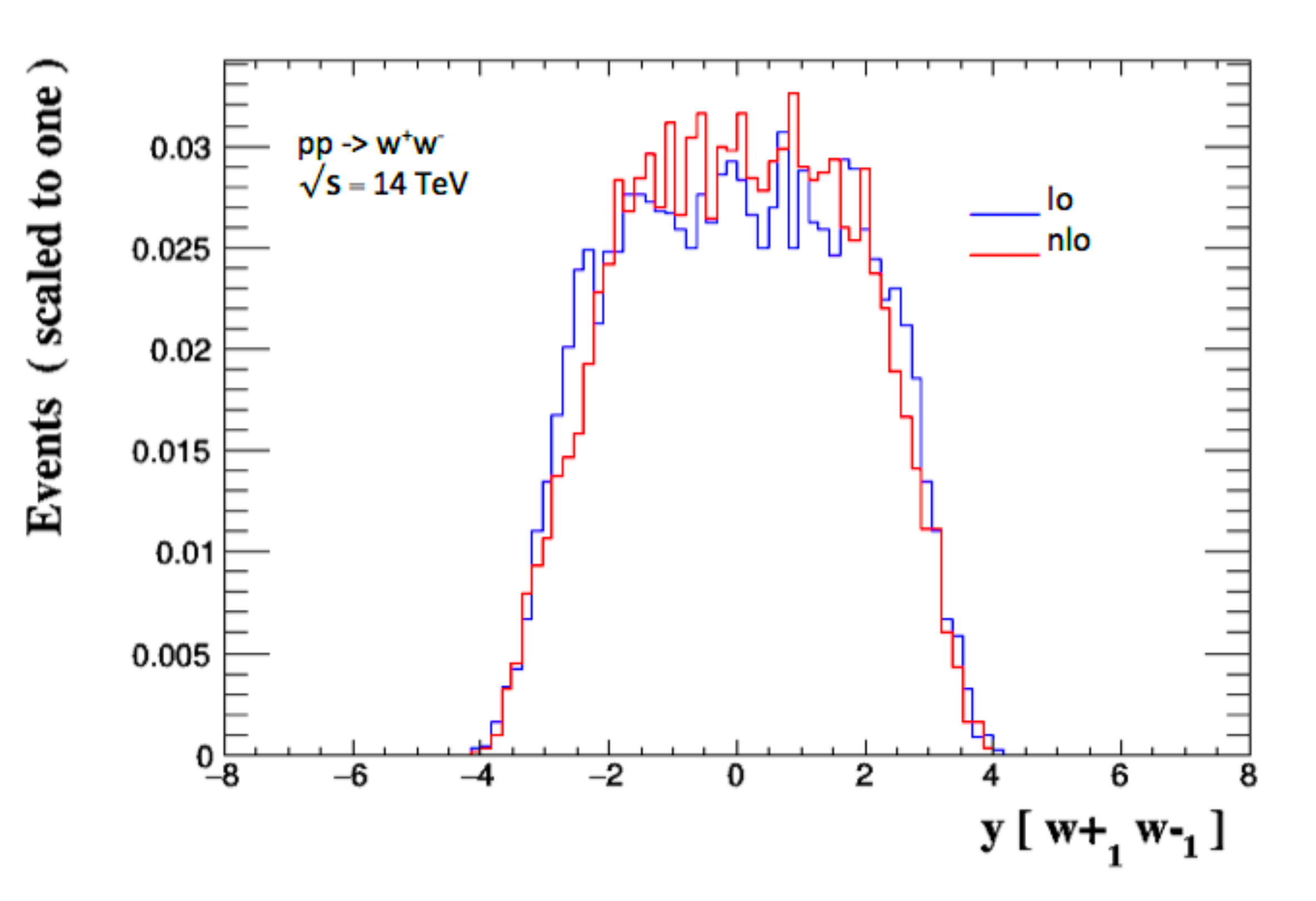} 	\includegraphics[width=.45\textwidth,trim=0 0 0 0,clip]{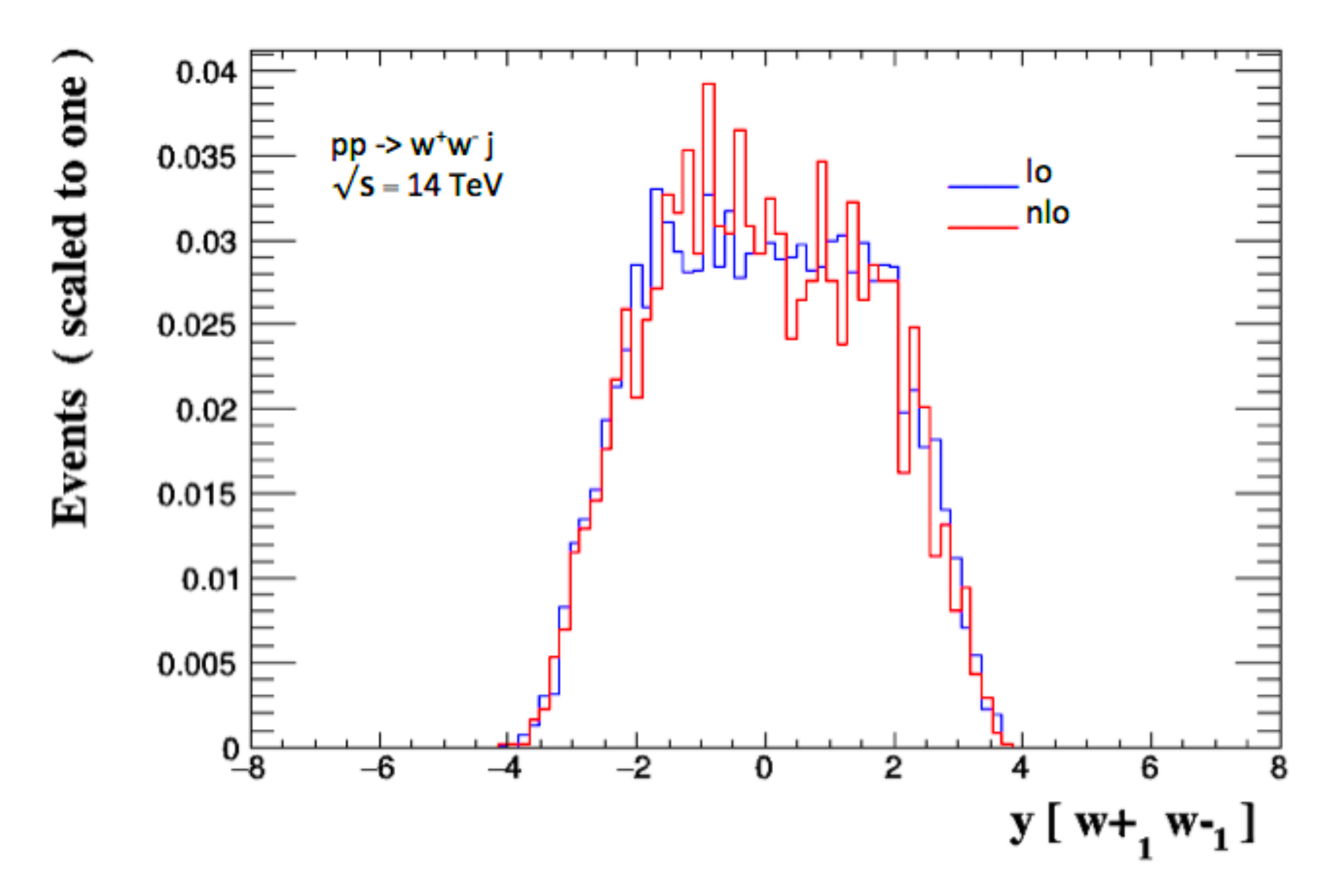} 
	\includegraphics[width=.45\textwidth,trim=0 0 0 0,clip]{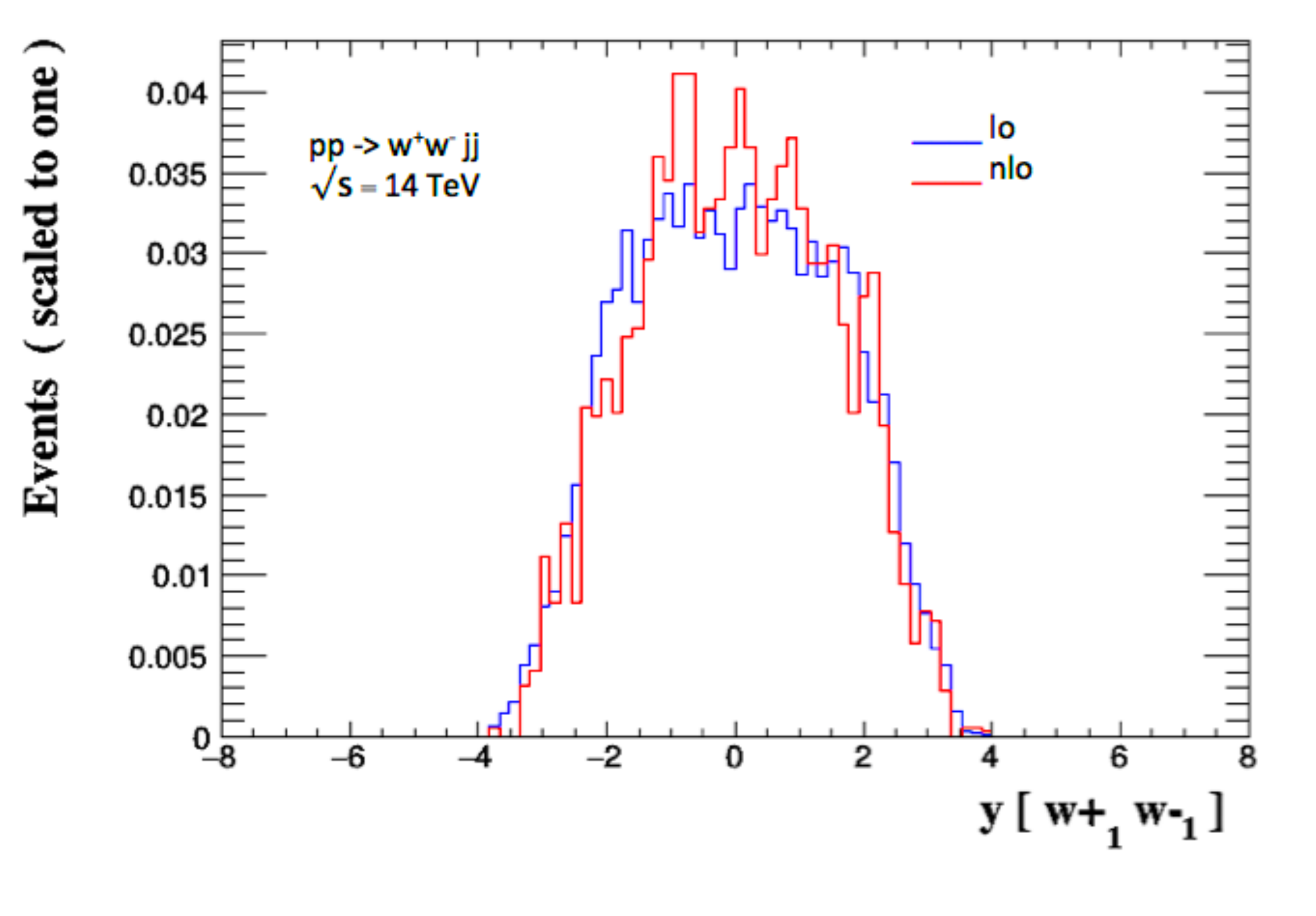} 
	
	\caption{ The distribution of rapidity $ y(W^{+ } W^{-})$ at partonic level for $W^{+}W^{-}$ production with 0, 1 and 2 jets at $\sqrt{s}= 14$ TeV
	}\label{fig:6}	
	
\end{figure}

For the processes $pp \rightarrow W^{+} W^{-} $, we observe that at low $P_{T}$ the LO and NLO contributions are extremely close to each other, as well in this region the  $p_{T}$ distribution is more interesting. However, at high values of $p_{T}$ the NLO contribution becomes more important. Although, the situation is different when we add one or two jets in the $W^{+} W^{-} $ pairs production final state , we observe a small QCD NLO correction effect on these distributions. 

We now turn to transverse momentum distribution of the $W^{+} W^{-} $. We observe  in Figure~\ref{fig:5} that the contributions for processes $pp \rightarrow W^{+} W^{-} j $ and $pp \rightarrow W^{+} W^{-} jj $ have the same form and the LO and NLO distributions are similar. We find that the process $pp \rightarrow W^{+} W^{-} $ to LO can not predict this variable it comes back to the existence of a single contribution that is $q \bar{q} $. Although at NLO the contributions included by the corrections of the QCD, in particular the high density of gluon at the small  $p_{T}$, show us another behaviour.

We give in figure~\ref{fig:6}  the distribution of rapidity $ y(W^{+ } W^{-})$. The three processes agree that the LO and NLO contributions are particularly sizeable near $-1<y(W^{+} W^{-}) <1 $, while the NLO correction are more considerable in the gap  $-1<y(W^{+} W^{-})<0 $.

\subsection{Parton Shower}
\label{subsec:4}

In this part, we will treat our processes at the hadron level, we have showered the events using Pythia8. Remembering that the reconstruction of events is done using Delphes fast detectors applying the ATLAS$\_$1604 07773.tcl card. 

\begin{table}[]
	\centering
	\begin{tabular}{|c|c|c|c|}
		\hline
		
		&	\multicolumn{3}{|c|}{$p^{min}_{T,j} > 20$ GeV} 	 \\ 	\hline 	
		&	$ \sigma_{LO+PS}$[pb]   &$\sigma_{NLO+PS}$[pb]   &$  K$    \\  \hline 
		
		$W^{+}W^{-}j$   &40,65 $ \pm$0.11 $_{-10.2\%}^{+11.7\%}$&  60.51 $ \pm$0.52 $_{-4.6\%}^{+4.0\%}$    & 1.45   \\ \hline 
		
		$W^{+}W^{-}jj$    & 23.32 $ \pm$0.057$_{-17.9\%}^{+23.9\%}$&   27.86 $\pm$ 0.17  $_{-5.0\%}^{+3\%}$&  1.19   \\ \hline	
		
	\end{tabular} 
	\caption{The LO and NLO total cross section for the various  processes of production at  $\sqrt{s}= 14$ TeV, with parton shower(PS) at hadronic level for different cuts on jet transverse momentum $P_{T,j} > 20$ GeV. }
	\label{tab:4}
	
	\centering
	\begin{tabular}{|c|c|c|c|}
		\hline
		
		&	\multicolumn{3}{|c|}{$p^{min}_{T,j} > 100$ GeV} 	 \\ 	\hline 	
		&	$ \sigma_{LO+PS}$[pb]   &$\sigma_{NLO+PS}$[pb]   &$  K$    \\  \hline 
		
		$W^{+}W^{-}j$   &	7.697$ \pm$0.018 $_{-10.2\%}^{+12.0\%}$ & 12.21 $ \pm$0.045 $_{-5.2\%}^{+4.6\%}$ & 1.57   \\ \hline 
		
		$W^{+}W^{-}jj$    & 1.965 $ \pm$0.0066 $_{-19.5\%}^{+26.6\%}$& 2.58 $ \pm$0.013$_{-7.5\%}^{+4.4\%}$ & 1.27   \\ \hline		
		
	\end{tabular} 
	\caption{The LO and NLO total cross section for the various  processes of production at  $\sqrt{s}= 14$ TeV, with parton shower(PS) at hadronic level for different cuts on jet transverse momentum  $p_{T,j} > 100$ GeV. }
	\label{tab:5}
\end{table}

First, We give the LO and NLO total cross sections of $W^{+} W^{-} $ pair production:

\begin{equation}
\label{eq:A}
\sigma_{LO+PS}[pb] = 66.76 \pm 0.17_{-6.65\%}^{+5.69\%}
\end{equation}
\begin{equation}
\label{eq:B}	
\sigma_{NLO+PS}[pb] = 115.114 \pm 0.4_{-4.4\%}^{+3.9\%},
\end{equation}
thus the K factor is equal to 1.67. We show in table~\ref{tab:4} and~\ref{tab:5} the measurements of LO and NLO cross sections of the processes $ p p  \rightarrow W^{+} W^{-}j $ and $ p p  \rightarrow W^{+} W^{-}jj $ at hadronic level. As we can see, the parton shower corrections preformed on our processes have a little weight on the measurements of the total cross section compared to  partonic results. In all these calculations, we obtained the theoretical  uncertainties due to PDFs sources between $\pm3.48\%$ and $5.59\%$. 

\begin{figure}[]
	\centering	
	
	\includegraphics[width=.45\textwidth,trim=0 0 0 0,clip]{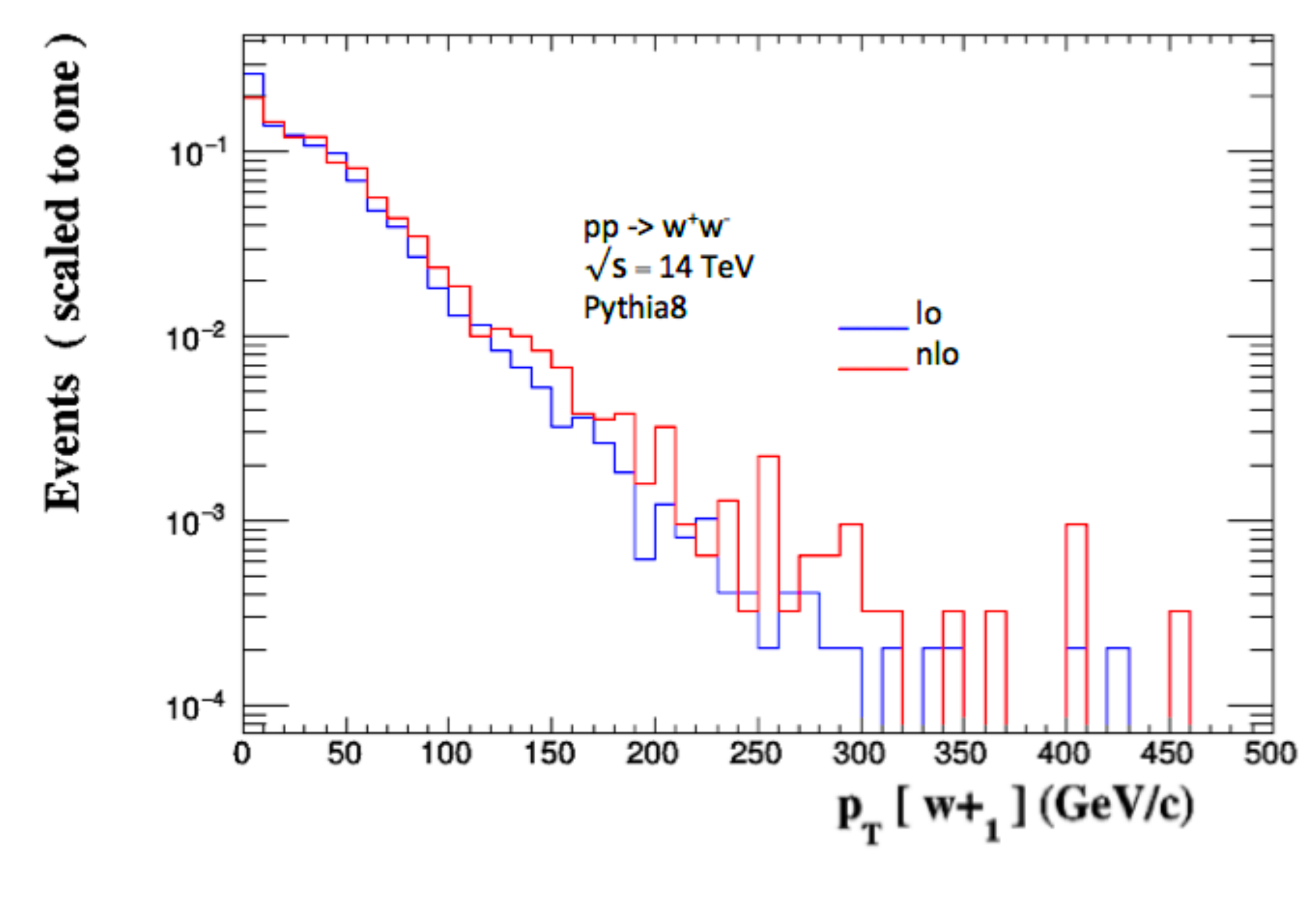} 	\includegraphics[width=.45\textwidth,trim=0 0 0 0,clip]{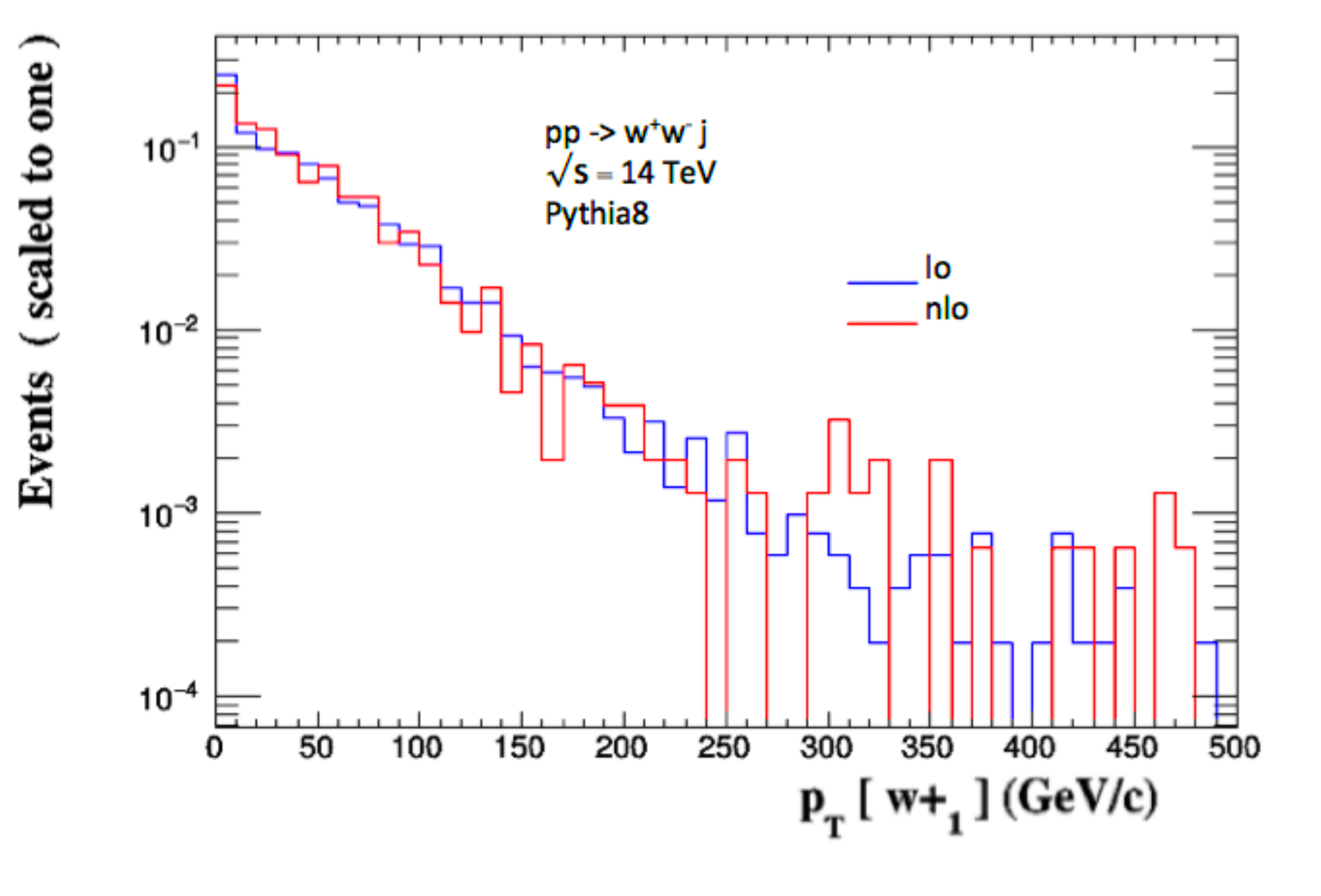} 
	\includegraphics[width=.45\textwidth,trim=0 0 0 0,clip]{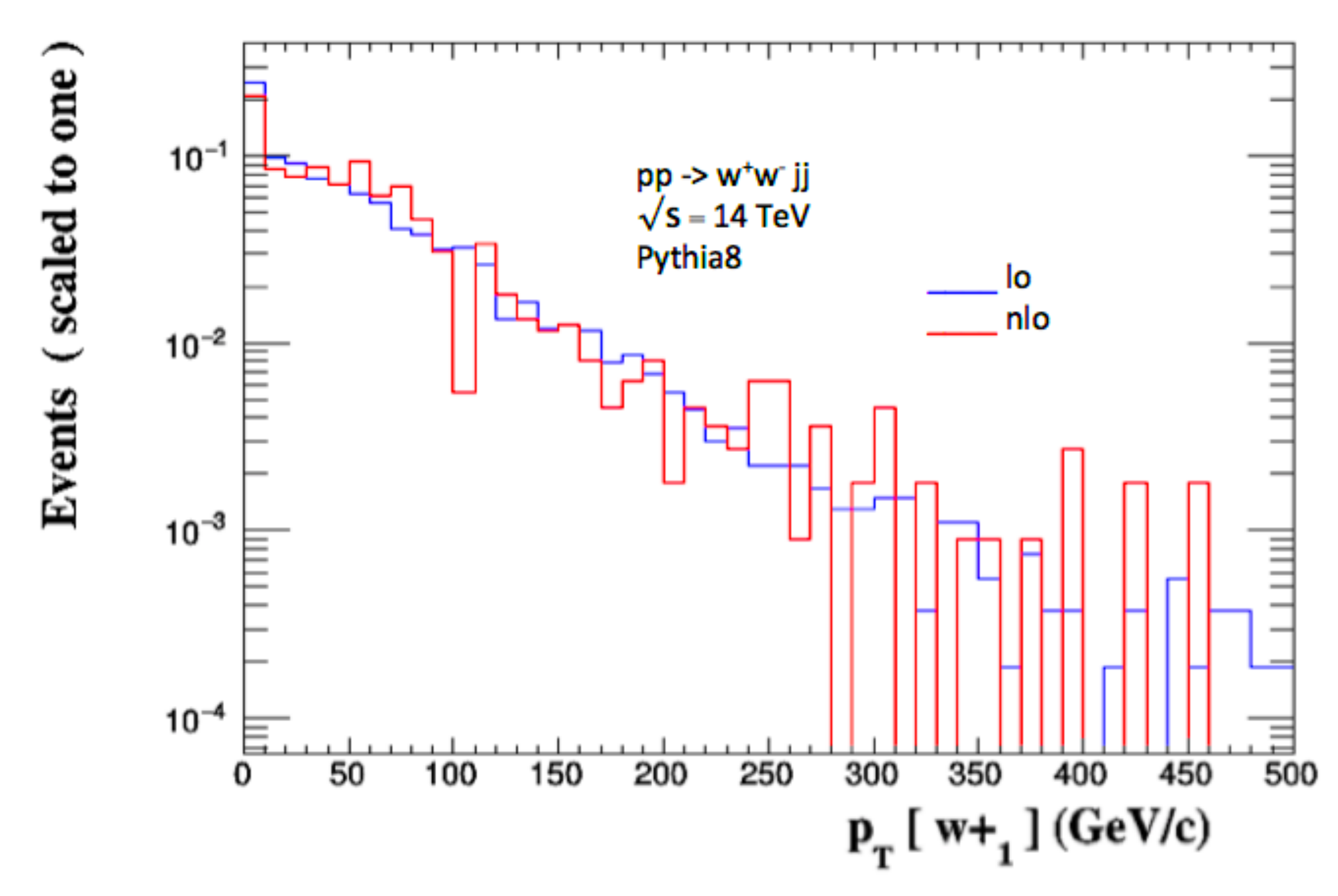} 
	
	\caption{ $W^{+} $ transverse momentum distribution at hadronic level for $W^{+}W^{-}$ production with 0, 1 and 2 jets at $\sqrt{s}= 14$ TeV.
	}\label{fig:7}	
\end{figure}

\begin{figure}[]
	\centering	
	\includegraphics[width=.45\textwidth,trim=0 0 0 0,clip]{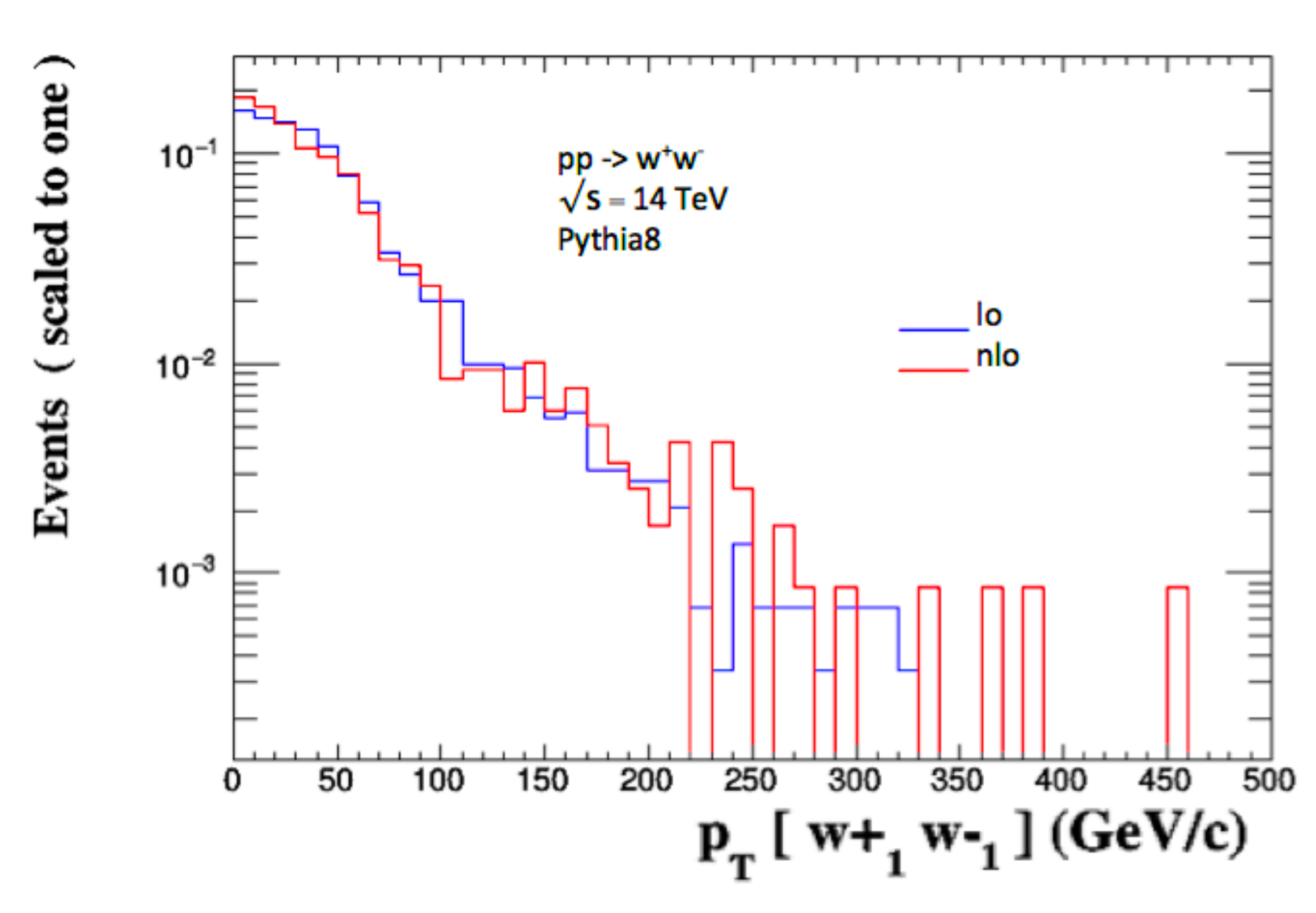} 
	\includegraphics[width=.45\textwidth,trim=0 0 0 0,clip]{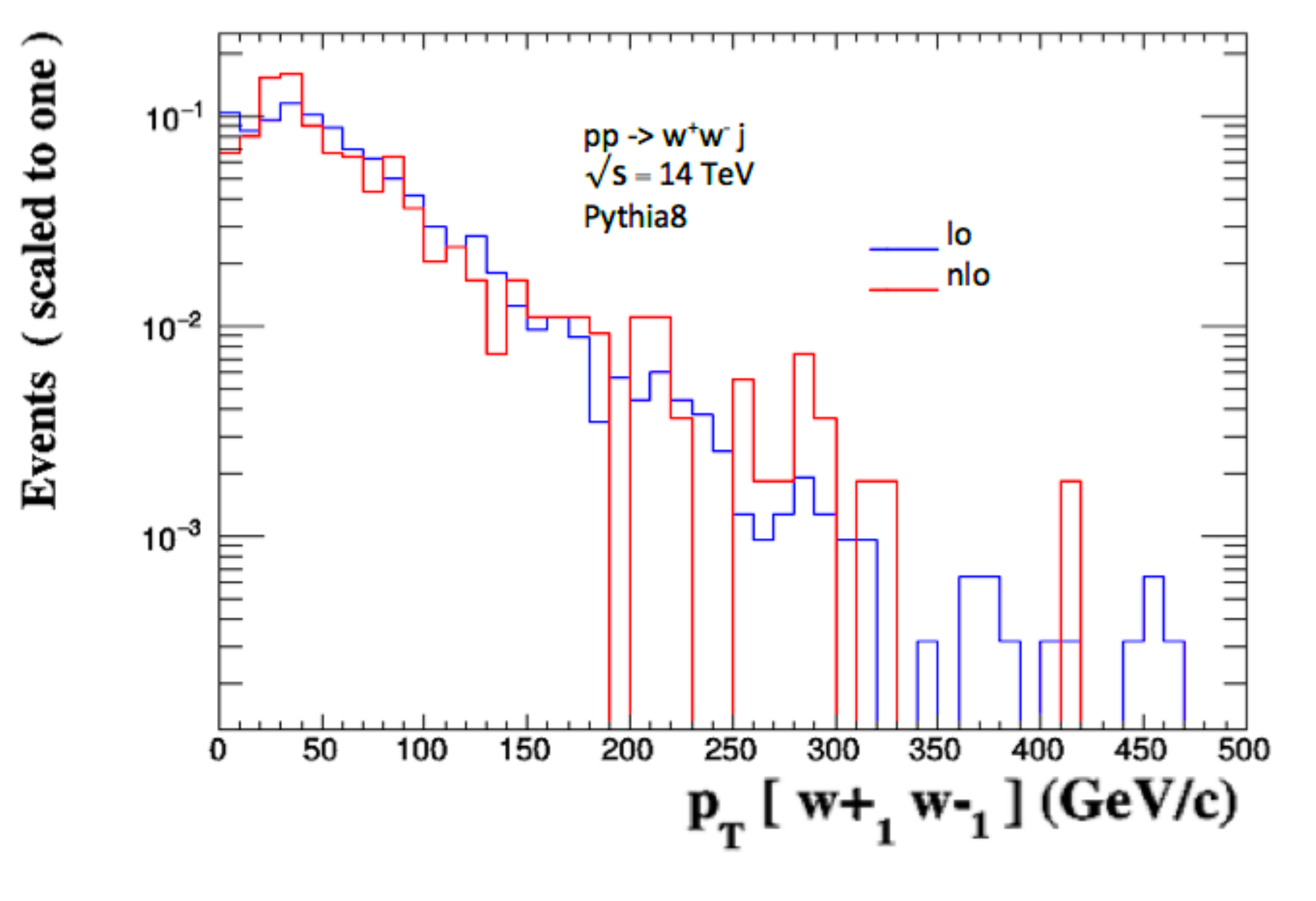} 
	\includegraphics[width=.45\textwidth,trim=0 0 0 0,clip]{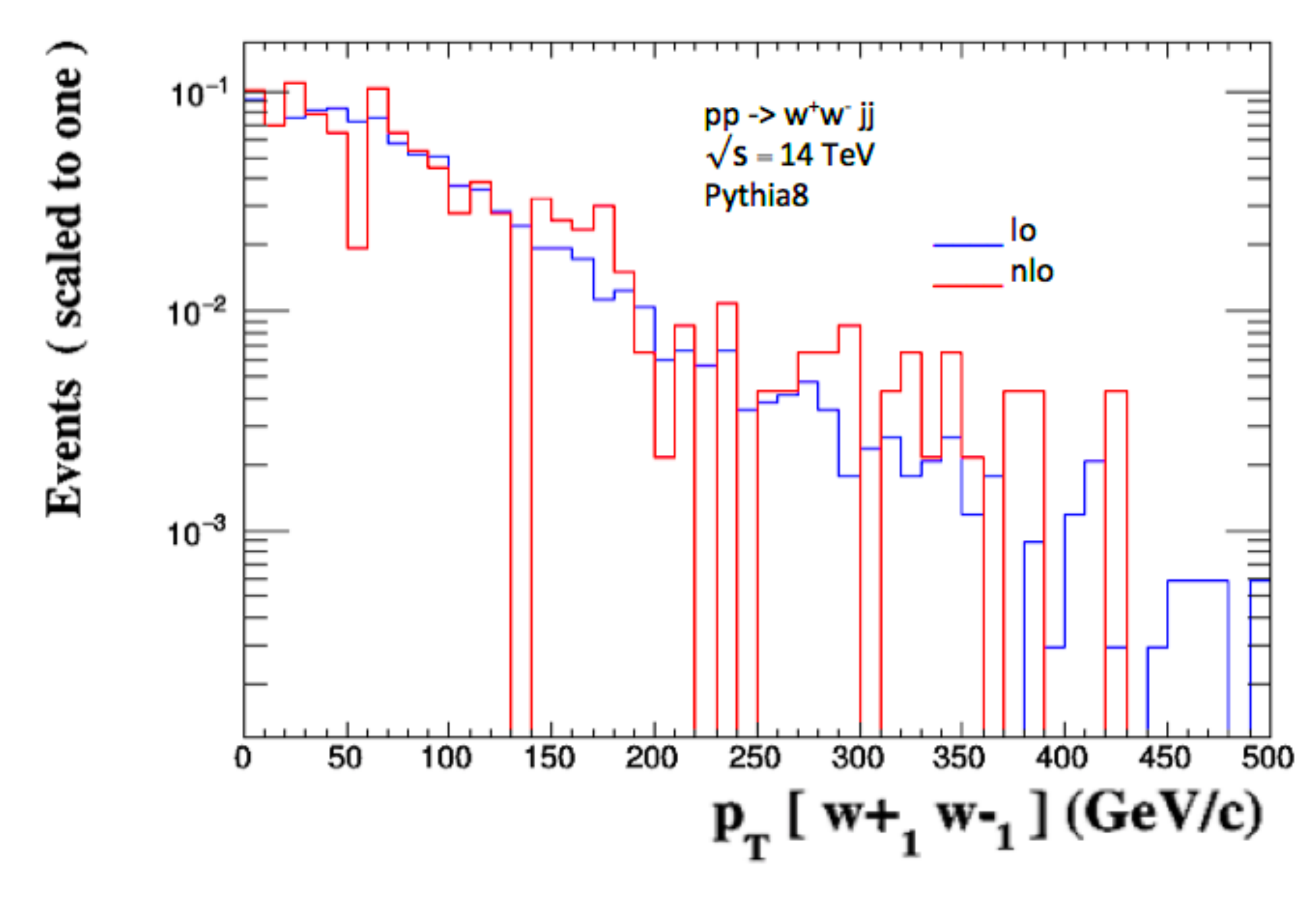} 
	
	\caption{  $W^{+}W^{-} $ transverse momentum distribution at hadronic level for $W^{+}W^{-}$ production with 0, 1 and 2 jets at $\sqrt{s}= 14$ TeV.
	}\label{fig:8}	
	
	\includegraphics[width=.45\textwidth,trim=0 0 0 0,clip]{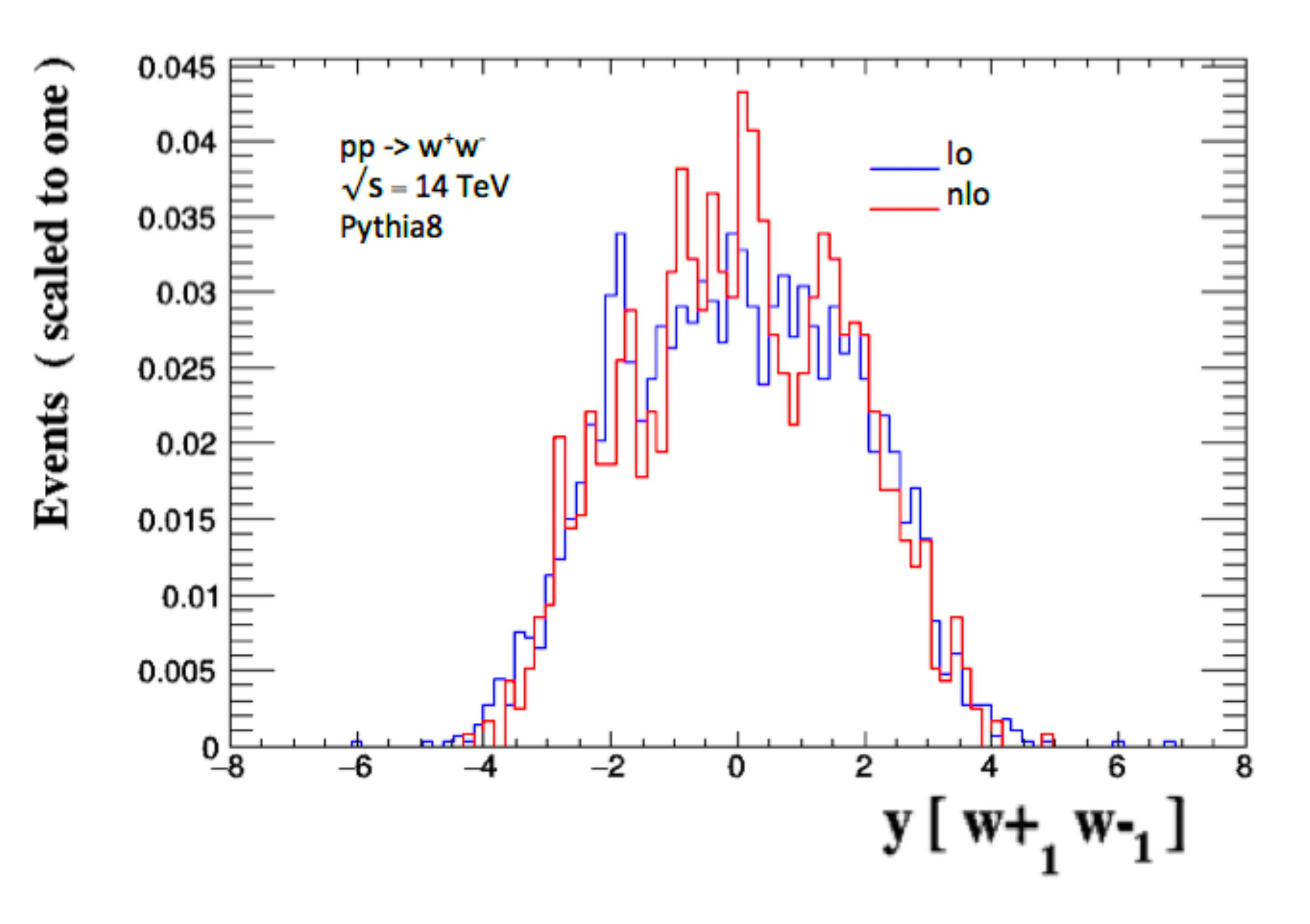} 	\includegraphics[width=.45\textwidth,trim=0 0 0 0,clip]{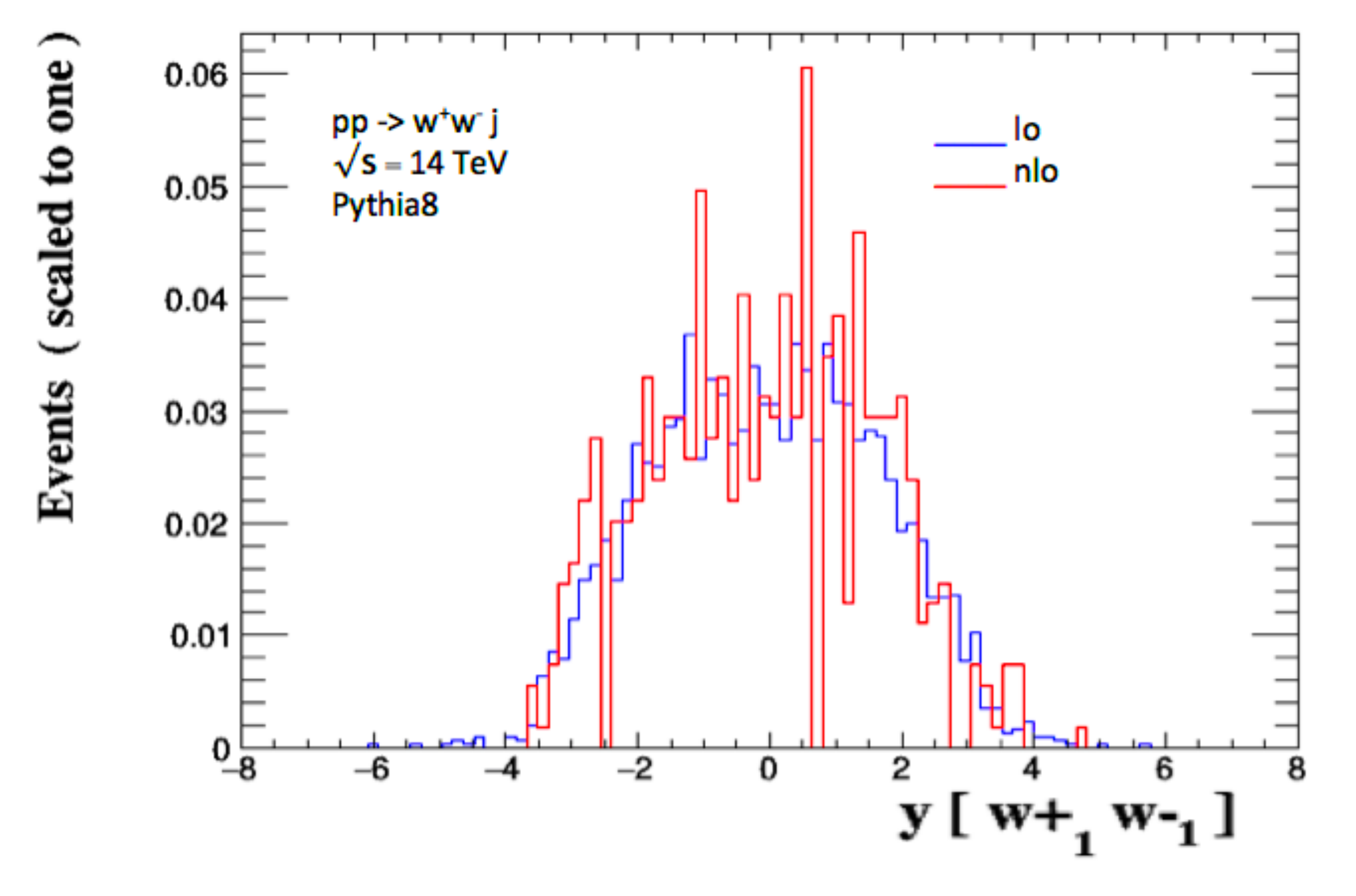} 
	\includegraphics[width=.45\textwidth,trim=0 0 0 0,clip]{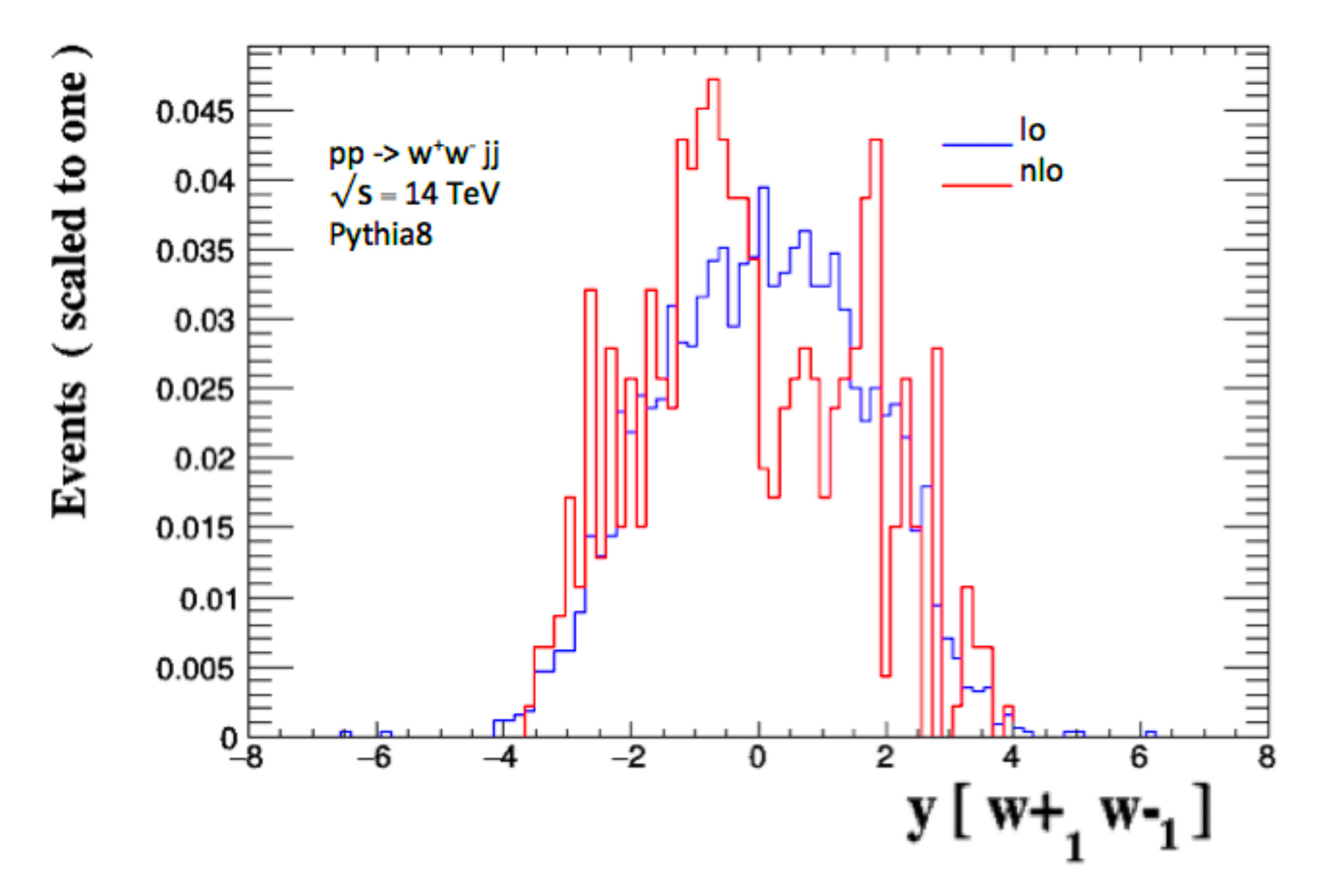} 
	
	\caption{  The distribution of rapidity $ y(W^{+ } W^{-})$ at hadronic level for $W^{+}W^{-}$ production with 0, 1 and 2 jets at $\sqrt{s}= 14$ TeV	}
	\label{fig:9}	
\end{figure}

We present in figure~\ref{fig:7}  the distributions of  $p_{T}(W^{+}) $  for all the processes $ p p  \rightarrow W^{+} W^{-} $ with 0, 1 and 2 jets and we note that here, we apply the cuts given in equations~\ref{eq:5} and~\ref{eq:6}. We observe that  the contributions have the same shape, they decrease on all the values considered, contrary to the partonic scale where two regions were found as shown in figure ~\ref{fig:3}. In the same figure~\ref{fig:7}, we see also that the $p_{T}(W^{+}) $  distribution for the $W^{+}W^{-} $ pair without jet, are more important in the $10$ GeV $<p_{T}< 160$ GeV region, but the QCD NLO corrections to this quantities are greater for the large $p_{T}$ ($p_{T}> 160$ GeV ). Although, these corrections change when adding a jet to the $p p  \rightarrow W^{+}W^{-} $  process in the final state.

We present in the figure~\ref{fig:8} the $W^{+}W^{-} $ transverse momentum distributions at LO and NLO. We notice that the distributions of the processes including one and two jets in the final state decrease rapidly compared to the process without jets. The QCD corrections are significant for $p_{T}(W^{+}W^{-} ) > 150 $ GeV. 

It is interesting to say that the $ p p  \rightarrow W^{+} W^{-}j $ and $ p p  \rightarrow W^{+} W^{-}jj $ processes, unlike the $ pp  \rightarrow W^{+} W^{-} $ process, have events at $p_{T}(W^{+}W^{-} ) > 300 $ GeV which means that an hypothetical signal of a new physics beyond the standard model can be detected at these transverse momentums.

We see in figure~\ref{fig:9}  the behaviour of the distribution of rapidity $ y(W^{+ } W^{-})$. We observe that the maximum interval is prolonged to $ -2 < y < 2$, but it is clear that adding a jet to these, necessarily changes the shape of this distribution.

\section{Conclusion}
\label{sec:5} 
In this paper, we present calculations for the production of $W^{+} W^{-}$ pairs of gauge bosons  associated to 0, 1 and 2 jets in proton-proton collisions at LHC at $\sqrt{s}$ = 14 TeV. Thus, we calculate their total cross sections at LO and NLO with QCD corrections with the help of MadGraph$\_$aMC@NLO using the NNPDF23 PDFs and the ATLAS$\_$1604$\_$07773.tcl card, which allow to have predictions greatly close to experimental results.

The NLO-QCD corrections have  a significant influence on the measurement of the total cross sections. The value of cross section at next-to-leading-order varies from 28 $\%$ - 74$\%$ of the ones at leading order. 

The results show that the LO and NLO total cross sections of the process $ p p  \rightarrow W^{+} W^{-}j$ and $ p p  \rightarrow W^{+} W^{-}jj$  have  the same behaviors with cuts in jet's transverse momentum $p_{T,j} > 20$ GeV and $p_{T,j} > 100$ GeV. We find that at large jets's transverse momentum $p_{T,j} > 100$ GeV, the LO and NLO total cross sections decrease. Contrary to the process $ p p  \rightarrow W^{+} W^{-}$ that has a jet QCD coming from the real contributions at NLO, the latter does not take into account our cuts. 

It has been shown that the appearance of the jets in the final state of production process of the $W^{+} W^{-}$ pairs decreases the total cross section. Indeed it is clear that the process $ p p  \rightarrow W^{+} W^{-}$ has the highest value of cross section.

The $W^{+}$ and $W^{+} W^{-}$ distributions at partonic level for our processes have the same appearance, of more we have distinguished two regions. A low growth up to $p_{T,j} = $ 60 GeV then a continuous decrease with $p_{T,j}$. At hadronic level with parton-shower effects, the distributions only decrease. 

At partonic level, the distribution of the rapidity of the pairs $y(W^{+} W^{-}) $ has a peak at $-1< y(W^{+} W^{-}) < 1$. With parton-shower, the interval is wider, $-2< y(W^{+} W^{-}) < 2$. The NLO QCD effects do not change these interval but, interestingly, they increase the peak values.

Finally, at higher transverse momentums ie. $p_{T} > 500$ GeV, the  yield is not-negligible. These processes can deliver potential background to several searches for physics beyond the SM.

\section*{Acknowledgement:} 
We thank Fawzi Boudjema for his careful reading and useful comments on the manuscript and for his enriching discussions. We thank him also for welcoming us at the internship in LAPTh (laboratoire d'Annecy-le-Vieux de Physique Th\'eorique).

\end{document}